\documentclass[%
reprint,
superscriptaddress,
amsmath,amssymb,
aps,
prb,
]{revtex4-2}

\usepackage{xcolor}

\usepackage{graphicx}% Include figure files
\usepackage{dcolumn}% Align table columns on decimal point
\usepackage{bm}% bold math
\usepackage{hyperref}% add hypertext capabilities
\begin{document}

\date{\today}

\title{Machine Learning and First-Principles Predictions of Materials with Low Lattice Thermal Conductivity}

\author{Chia-Min Lin}
\affiliation{Department of Physics, University of Alabama at Birmingham, Birmingham, Alabama 35294, USA}

\author{Abishek Khatri}
\affiliation{Department of Physics, University of Alabama at Birmingham, Birmingham, Alabama 35294, USA}

\author{Da Yan}
\affiliation{Department of Computer Sciences, Indiana University Bloomington, Bloomington, Indiana 47405, USA}

\author{Cheng-Chien Chen}
\email{chencc@uab.edu}
\affiliation{Department of Physics, University of Alabama at Birmingham, Birmingham, Alabama 35294, USA}

\begin{abstract}
We perform machine learning (ML) simulations and density functional theory (DFT) calculations to search for materials with low lattice thermal conductivity, $\kappa_L$. Several cadmium (Cd) compounds containing elements from the alkali-metal and carbon groups including A$_2$CdX (A = Li, Na, and K; X = Pb, Sn, and Ge) are predicted by our ML models to exhibit very low $\kappa_L$ values ($< 1.0 $ W/mK), rendering these materials suitable for potential thermal management and insulation applications. Further DFT calculations of electronic and transport properties indicate that the figure of merit, $ZT$, for thermoelectric performance can exceed 1.0 in compounds such as K$_2$CdPb, K$_2$CdSn, and K$_2$CdGe, which are thereby also promising thermoelectric materials.
\end{abstract}

\keywords{lattice thermal conductivity, thermoelectric material, machine learning, density functional theory}

\maketitle

\section{Introduction}

Materials with low lattice thermal conductivity ($\kappa_L$) have important applications in thermal management and energy conversion, by serving as thermal insulation and barrier coatings, or as thermoelectrics. In particular, thermoelectric (TE) materials can directly convert between thermal and electrical energy~\cite{he2017advances, liu2017new, zevalkink2018practical, urban2019new, wei2020review,hasan2020inorganic,zoui2020review, jaziri2020comprehensive}, offering potential solutions for sustainable clean energy. To date, however, large-scale TE applications remain limited due to the relatively low energy conversion efficiency of known materials. Improving the efficiency and finding suitable TE materials that function at different temperatures remain important tasks in materials science research.

The TE performance can be quantified by the figure of merit value $ZT = S^2 \sigma T/\kappa$, where $S$ is the Seebeck coefficient (the induced voltage in response to a temperature gradient), $\sigma$ is the electrical conductivity, $T$ is the temperature, and $\kappa$ is the thermal conductivity. One approach to enhance $ZT$ is by increasing the power factor ($S^2 \sigma$), via band engineering of carrier concentration and mobility, among other factors~\cite{pei2014high,lee2015enhanced,lu2015increasing, jiang2021entropy, ma2021review,ghosh2022insights,ding2023xmosin2}. Another approach is to find materials with low thermal conductivity~\cite{zhu2021charting,loftis2020lattice,tranaas2022lattice}.

There are two major contributions to thermal conductivity: $\kappa = \kappa_e + \kappa_L$. In general, the electronic contribution $\kappa_e$ closely follows the Wiedemann-Franz law~\cite{chester1961law}, $\kappa_e = L \sigma T$, where $L$ is the Lorenz number ($2.44 \times 10^{-8}$ W$\Omega$/K$^2$ for free electrons). $\kappa_e$ also varies with the charge carrier concentration $n$. On the other hand, $\kappa_L$ has a distinct $T$ dependence. If the lattice contribution $\kappa_L$ of a material is much lower than the electronic contribution $\kappa_e$ under certain $n$
and $T$ conditions, an optimal $ZT \sim S^2/L > 1$ can be achieved due to the Wiedemann-Franz law.
Therefore, designing or searching for materials with low $\kappa_L$ continues to be an active research area, employing approaches such as phonon engineering, nanostructuring, and/or applying external strain or pressure~\cite{seko2015prediction, yang2021high, lin2021first, wu2022enhanced,govindaraj2022pressure,qi2022large, xia2024strain}.

Computational materials modeling has played an important role in providing predictions and critical insights into the thermal conducting behavior of materials~\cite{gorai2017computationally, puligheddu2019computational,xia2020high, he2022accelerated,xia2023unified,ma2024multilayer}. Traditionally, density functional theory (DFT) is the standard computational workforce for accurate calculations of $\kappa_L$ from first principles. However, its relatively high computational cost limits large-scale investigations of $\kappa_L$ in new materials.
More recently, data-driven machine learning (ML) approaches have become popular and powerful tools for materials modeling and discovery~\cite{gaultois2013data, furmanchuk2018prediction, choudhary2020data, wang2020machine, liu2020high, chen2021machine, mbaye2021data, han2021machine,american2022machine, wang2023critical, wu2023machine, purcell2023accelerating,liu2024mathub}. 
This popularity and improvement in ML research largely result from advancements in computer architectures and ML algorithms, as well as from the increasing availability of open materials databases.
ML algorithms can learn from training data by identifying connections through linear or non-linear relationships between target properties and input features.
Once trained, ML models can achieve highly efficient and often accurate large-scale predictions.
 
In this study, we utilize combined machine learning (ML) predictions and density functional theory (DFT) calculations to discover materials with low lattice thermal conductivity.
Specifically, we develop ensemble-tree ML models with input features based on the chemical formula and atomic configurations to quickly estimate $\kappa_L$ of a given material. For promising low-$\kappa$ materials identified by our ML models, we further validate the results by performing DFT calculations to evaluate $\kappa_L$ directly from first principles. In particular, we find that the chemical compositions A$_2$CdX (A = Li, Na, and K; X = Pb, Sn, and Ge) of orthorhombic crystal symmetry exhibit \textit{ultra-low lattice thermal conductivity} ($\kappa_L \sim 0.1-1.0$ W/mK). 
Our DFT calculations of the transport and thermoelectric properties further indicate that some of these materials like K$_2$CdPb can exhibit a $ZT\ge 1.0$ near room temperature, which are thereby promising for \textit{low-temperature thermoelectric application}~\cite{soleimani2020review}.

The rest of the paper is organized as follows: Section \ref{sec:methods} presents the computational details of machine learning (ML) models and first-principles density functional theory (DFT) calculations. Section \ref{sec:results} presents the ML and DFT predictions of low-$\kappa_L$ materials and their thermoelectric properties. Finally, Section \ref{sec:conclusion} concludes the paper by summarizing our main findings.

%%%%%%%%%%%%%%%%%%%%%%%%%%%%%%%%%%%%%%%%%%
\section{\label{sec:methods} Computational Methods}

\subsection{Machine Learning Simulation}
{\it Data Acquisition and Feature Creation --}
Our machine learning (ML) models aim at predicting the target property of lattice thermal conductivity $\kappa_L$ for a given compound. The ML training dataset is sourced from the TE Design Lab, which is a virtual platform hosting a database of calculated thermoelectric properties~\cite{gorai2016te}. From this database, we select a total of 1900 compounds with $\kappa_L$ in the range of $0 - 1100$ W/mK.
For all compounds in the selected dataset, we then use the Matminer package~\cite{ward2018matminer} to generate 61 input features based on their chemical formula and atomic configurations~\cite{chen2019machine}. These features can be broadly categorized as structural features and elemental features.
Specifically, 7 structural features include the space group, volume per atom, packing fraction, unit-cell density, bond length, bond angle, and cohesive energy. 
Moreover, 18 elemental features include the atomic mass, atomic radius, atomic number, periodic table group, row number, block number, Mendeleev number, molar volume, boiling point, melting temperature, Pauling electronegativity, first ionization energy, covalent radius, and volume per atom from ground state, as well as the average number of $s$, $p$, $d$, and $f$ valence electrons. 
Since our dataset contains compounds ranging from unary to quinary materials, each elemental feature can be expanded by calculating the minimum, maximum, and weighted average of the constituent chemicals, resulting in a total of 54 $(= 18 \times 3)$ elemental features. Overall, 61 $(= 7 + 54)$ features are used in training the ML models.

We note that several features, such as average atomic mass and volume (which is related to atomic radius), are relevant parameters for estimating $\kappa_L$ in known empirical formulas~\cite{callaway1959model,slack1973nonmetallic}. There, $\kappa_L$ is also expected to be proportional to the mean sound velocity $v_m$ (or the Debye temperature $\Theta_D$) cubed. It was shown that ML models can accurately predict $v_m$ and $\Theta_D$ using features simply derived from chemical compositions and crystal symmetry~\cite{smith2023machine}. Therefore, it is anticipated that ML models trained here with the 61 features under study can perform well in predicting $\kappa_L$~\cite{chen2019machine}.

{\it Model Training and Validation --}
Our supervised ML tasks utilize Random Forests as the underlying algorithm~\cite{ho1998random, breiman2001random}. Random Forest is an ensemble method consisting of multiple decision trees. Each tree is trained on a randomly selected subset of features and samples. The Random Forest algorithm then averages the results of all trees to make the final prediction, which generally reduces the overfitting problem associated with a single decision tree. Random Forest ML models are relatively easy to train and often produce highly accurate results. To further reduce overfitting, we also pre-prune the trees by limiting their depth. 
%Based on the GridSearchCV results, we set the maximum depth to 13 layers with 100 estimators, which is the best condition for a good balance between bias and variance. Deeper trees may lead to overfitting and will not improve model performance.
Specifically, we use 90\% of our input data as the training-validation set and apply the GridSearchCV technique from the scikit-learn library~\cite{pedregosa2011scikit} to determine the optimal tree depth via 10-fold cross-validation. The remaining 10\% of the input data serves as the unbiased test set to evaluate the final model performance. After training and evaluation, we then use the ML model to predict the lattice thermal conductivity $\kappa_L$.

\subsection{First-Principles Calculation}
For promising low-$\kappa_L$ materials identified by ML models, we further perform first-principles density functional theory (DFT) calculations to validate their thermoelectric properties.
Our calculations are based on the Vienna Ab initio Simulation Package (VASP, version 5.4.4)~\cite{kresse1996efficiency,kresse1996efficient}, which is a highly efficient and accurate plane-wave pseudopotential DFT code.
We adopt the projector-augmented-wave (PAW) potentials~\cite{blochl1994projector,kresse1999ultrasoft} and utilize the Perdew-Burke-Ernzerhof generalized gradient approximation (GGA-PBE) functional~\cite{perdew1996generalized}. 
The plane-wave cutoff energy is set to 500 eV, and a fine $\Gamma$-centered Monkhorst-Pack grid of $19 \times 19 \times 19$ points is used for Brillouin zone integration~\cite{monkhorst1976special}. 
For a given crystal structure, we first fully relax the lattice parameters and atomic positions.
The convergence criteria for the electronic and ionic relaxation loops are set to $10^{-8}$ eV per unit cell and $10^{-4}$ eV/{\AA}, respectively.

After structure relaxation, we compute the thermoelectric properties ($S$, $\sigma$, and $\kappa_e$) using the BoltzTraP2 package~\cite{madsen2018boltztrap2}, which is based on Boltzmann transport theory with a constant relaxation time approximation. The lattice thermal conductivity ($\kappa_{L}$) is obtained through first-principles phonon calculations using the Phonopy and Phono3py~\cite{phonopy,phono3py} packages, which are based on finite-displacement supercell approaches. Phonopy computes the phonon spectra at the harmonic or quasi-harmonic level. Phono3py evaluates phonon-phonon interactions and $\kappa_{L}$ from the Peierls-Boltzmann equation~\cite{phonopy-phono3py-JPCM}. In the supercell calculations, the atomic displacement is set to 0.02 {\AA}, and the real-space interaction cutoff distance is set to 4.0 {\AA}. For the second-order (harmonic) and third-order (anharmonic) phonon calculations, $3 \times 3 \times 3$ supercells with a $5 \times 5 \times 5$ k-mesh and $2 \times 2 \times 2$ supercells with a $9 \times 9 \times 9$ k-mesh are employed, respectively. A phonon q-point sampling mesh of $21 \times 21 \times 21$ points is used.
The theoretical crystal structure in this study is visualized using the VESTA software (version 3.4.8)~\cite{momma2011vesta}.

%%%%%%%%%%%%%%%%%%%%%%%%%%%%%%%%%%%%%%%%%%
\section{\label{sec:results}Results and Discussion}

\begin{figure*}
\includegraphics[width=1.0\linewidth]{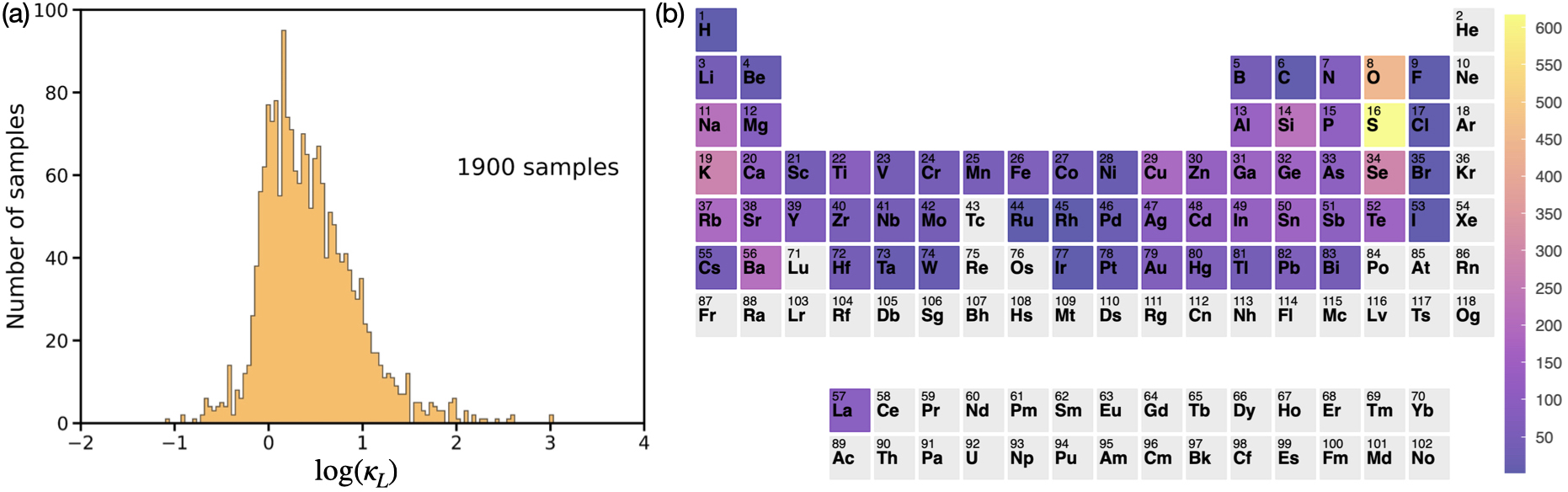}
\caption{(\textbf{a}) Histogram for the 1900 training samples of lattice thermal conductivity ($\kappa_{L}$) selected from the TE Design Lab~\cite{gorai2016te}. The distribution spans nearly five orders of magnitude and is plotted on a base-10 logarithmic scale. (\textbf{b}) False-color intensity plot showing the frequency of each element in the 1900 training dataset. Elements not present in the list are shown in gray. The figure was created using the open-source software Periodic Trend Plotter~\cite{PeriodicTrendPlotter}.
\label{fig1}}
\end{figure*} 

Figure \ref{fig1}(a) shows the distribution of $\kappa_{L}$ for the 1900 compounds in our training dataset from the TE Design Lab~\cite{gorai2016te}. Since the range of the distribution spans nearly 5 orders of magnitude, it is plotted on a base-10 logarithmic scale. Eventually, ML models are trained to predict $\log (\kappa_L)$.
For the accuracy and generalizability of our ML models, we ensure that our dataset is diverse in chemical composition (from unary to quinary compounds) and crystal structure (with 140 different space groups). 
In particular, among the 1900 samples, 7 are unary, 418 are binary, 1143 are ternary, 328 are quaternary, and 4 are quinary. These compounds contain 61 different atomic elements; the frequency each element appears in the compound list is represented by the false-color intensity plot in Figure \ref{fig1}(b). Gray color means that the element is not present.

\begin{figure}
\includegraphics[width=1.0\linewidth]{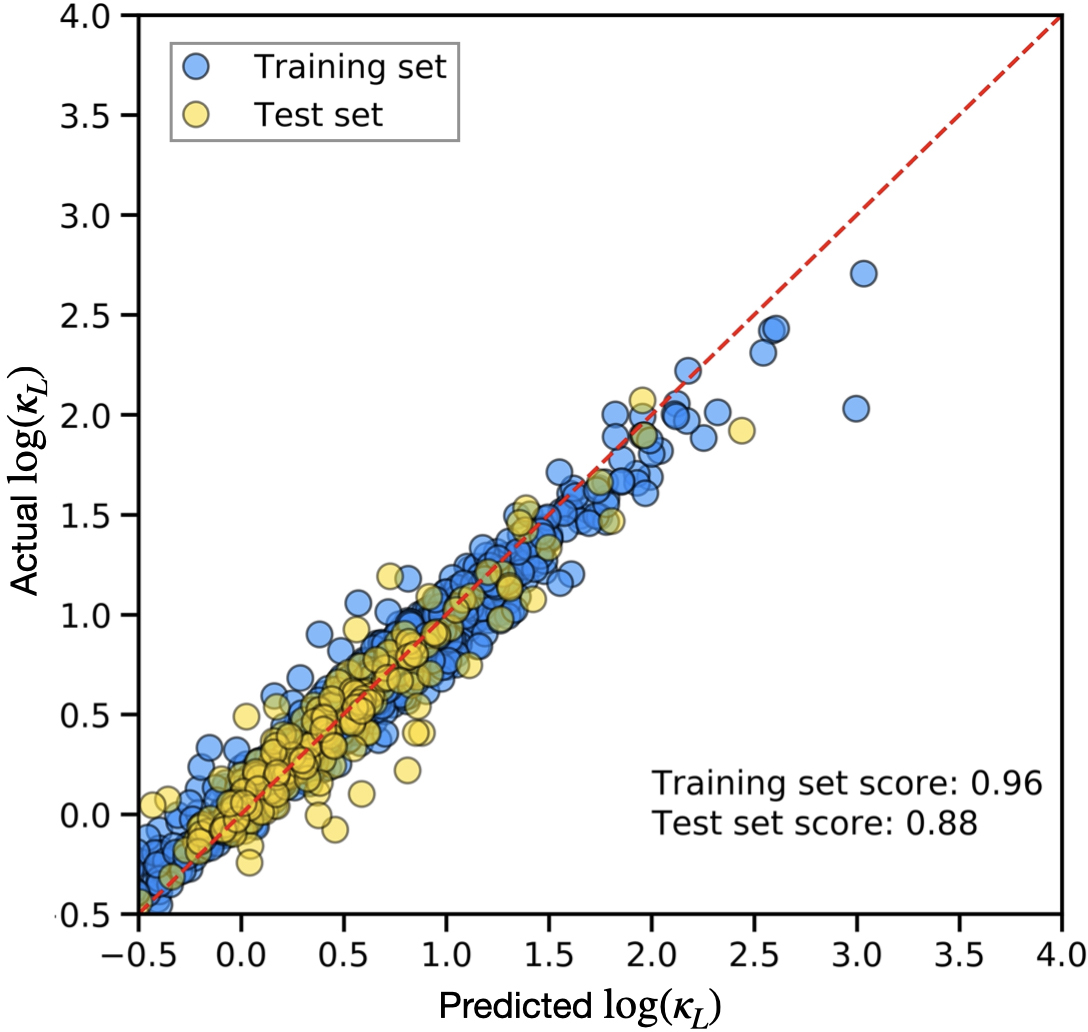}
\caption{Evaluation of the Random Forests model in predicting the logarithmic value of lattice thermal conductivity, $\log(\kappa_{L})$. The training and test sets consist of 90\% and 10\% of our total dataset (1900 samples), respectively. When the machine learning prediction perfectly matches the actual value, the data point will fall on the red dashed line. The model achieves relatively high $R^2$ scores of 0.96 and 0.88 for the training and test sets, respectively.
\label{fig2}}
\end{figure}

As discussed in Section \ref{sec:methods}, our ML models are based on Random Forests trained with 61 features~\cite{chen2019machine} generated by the Matminer package~\cite{ward2018matminer}.
The coefficient of determination $R^{2}$ is used to evaluate the model performance:
\begin{eqnarray}
 R^{2}=1-\frac{\sum_i(y_{i}-\hat{y_{i}})^{2}}{\sum_i(y_{i}-\overline{y})^{2}},
\end{eqnarray}
where $y_{i}$, $\hat{y_{i}}$, $\overline{y}$ are the actual value (for the $i$-th entry), the predicted value, and the mean of the actual values, respectively. $R^{2}$ ranges from 0 to 1, with $R^2 = 1$ indicating a perfect prediction.
Figure \ref{fig2} shows the resulting ML model performance on predicting $\log (\kappa_L)$. The blue and yellow circles represent data from the training-validation set (90\%) and the test set (10\%), respectively. A red dashed line is also plotted as a guide to the ideal line where the predicted values match the actual values. Our model achieves an $R^{2} =0.96$ for the training-validation set and $R^2 =0.88$ for the test set, indicating that our ML model provides a fairly accurate prediction of $\log(\kappa_L)$.

Random Forest models also provide information on feature importance in the ML predictions. Among the features under study, the atomic bond length is found to be the most significant factor affecting $\kappa_L$.
In an ideal gas model, lattice thermal conductivity is approximated as
\begin{equation}
\kappa_L = \frac{1}{3} v^2_s c_v \tau_s,
\end{equation}
where $v_s$ is the phonon velocity, $c_v$ is the specific heat, and $\tau_s$ is the phonon relaxation time. 
Among these three parameters, the phonon relaxation time is related to bond-strength anharmonicity \cite{dutta2021evidence,chen2018manipulation,chang2018anharmoncity}, which is correlated with bond length.
In particular, a longer bond length is prone to causing anharmonic vibrations, as the interatomic force constant decreases with increasing bond length. Anharmonicity then facilitates collisions between different phonon modes. As anharmonicity increases, the phonon relaxation time decreases, which in turn leads to a reduction in lattice thermal conductivity.

We note that the strength of anharmonicity also can be evaluated by the Gr\"{u}neisen parameter:
\begin{equation}
\gamma = \frac{V}{\omega}\frac{\partial\omega}{\partial V},
\end{equation}
where $V$ is the crystal volume and $\omega$ is the phonon frequency. 
Within the harmonic approximation, the thermal expansion is zero on average. In the presence of anharmonicity, the phonon frequency can vary as the volume changes with temperature. Therefore, a larger Gr\"{u}neisen parameter indicates stronger anharmonicity and a lower lattice thermal conductivity.
In fact, based on the Debye-Callaway model~\cite{callaway1959model,slack1973nonmetallic}, the lattice thermal conductivity can be approximately evaluated as
\begin{equation}
\kappa_{L} \approx \frac{Mv^{3}_{m}}{TV^{2/3} \gamma^{2}}\frac{1}{N^{1/3}},
\end{equation}
where $M$, $v_{m}$, $T$, $V$, $\gamma$, and $N$ represent the average mass, the mean speed of sound, the temperature, the average volume per atom, the Gr\"{u}neisen parameter, and the number of atoms per primitive unit cell, respectively. The above formula shows that $\kappa_L$ is inversely proportional to $\gamma^2$ and $V^{2/3}$.
Indeed, in addition to bond length, the volume per atom is evaluated by our ML models as the second most important feature affecting $\kappa_L$. Overall, the feature importances align well with the above approximated models for $\kappa_L$, demonstrating that our ML models are reasonable and adequate.

\begin{figure}
\includegraphics[width=1.0\linewidth]{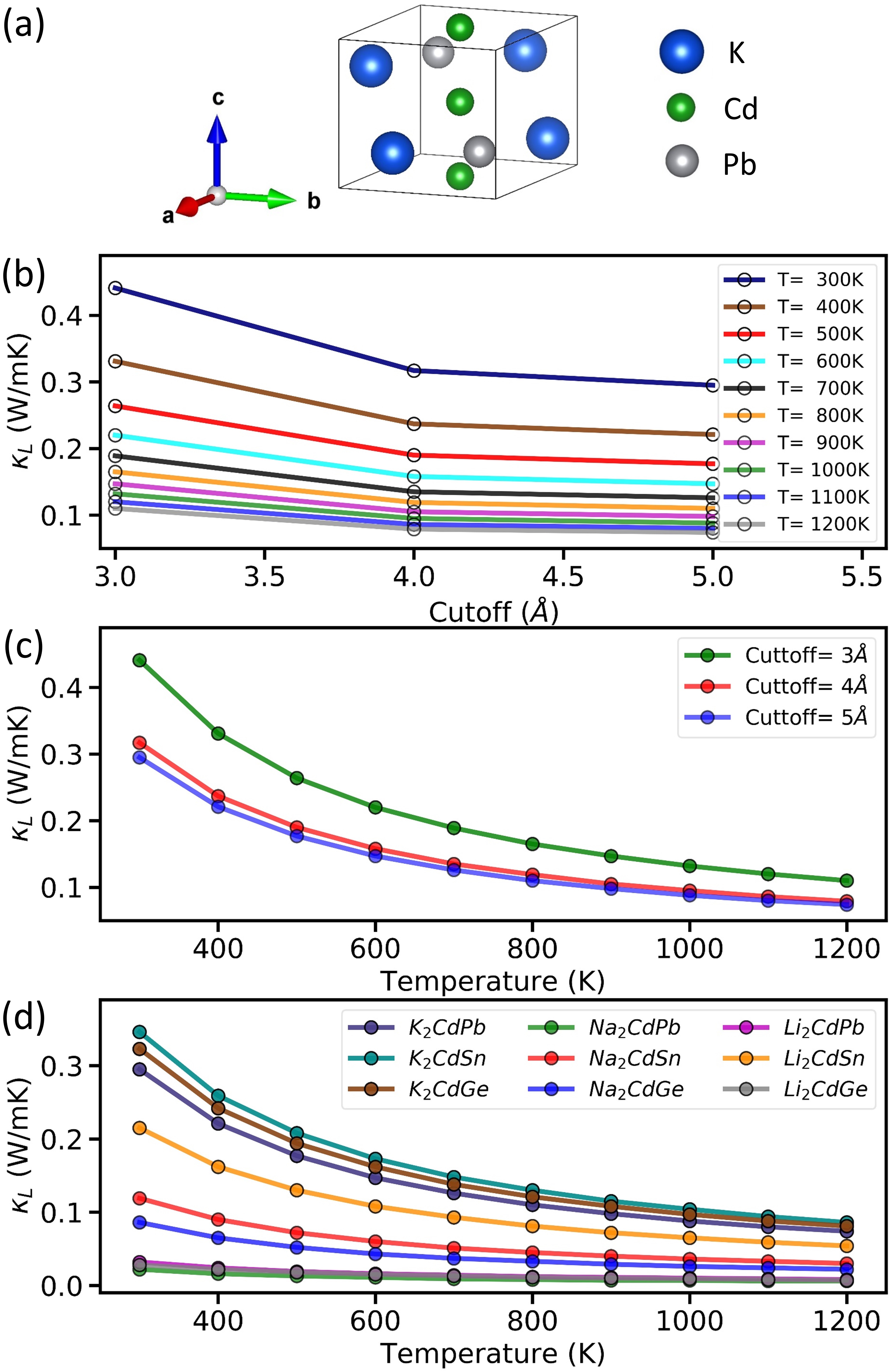}
\caption{(\textbf{a}) Primitive-cell crystal structure of K$_2$CdPb, with orthorhombic symmetry and space group Ama2 (No. 40). (\textbf{a}) Lattice thermal conductivity $\kappa_{L}$ of K$_2$CdPb as a function of the neighbor interaction cutoff distance in the temperature range of $300 - 1200$ K. (\textbf{a}) $\kappa_{L}$ of K$_2$CdPb as a function of temperature for different cutoff distances. (d) $\kappa_{L}$ computed with a cutoff distance of 4 {\AA} for nine different Cd-compounds.
\label{fig3}}
\end{figure}

We next apply the ML models to predict materials with low $\kappa_L$. Recently, Zintl-phase compounds have attracted significant attention due to their strong anharmonic properties, which could lead to low lattice thermal conductivities~\cite{toberer2010zintl,zevalkink2012thermoelectric, ding2018low, yin2019review,cai2019promising, guo2021unveiling, wang2023acoustic, tranaas2023lattice}. The Zintl phase refers to compounds formed by alkali metals (group I) or alkaline-earth metals (group II) combined with $p$-block metals or metalloids (from groups III-VI).
Other recent studies have also shown that Zintl-phase compounds can achieve ultra-low $\kappa_L$ by introducing a heavy element, cadmium (Cd)~\cite{pandey2015high,seko2015prediction}. For these reasons, we focus on applying our ML models to Cd-based Zintl-phase materials. Specifically, we consider A$_2$CdX (A = Li, Na, and K; X = Pb, Sn, and Ge) with orthorhombic symmetry and space group Ama2 (No. 40)~\cite{zhang2022remarkable}; Figure \ref{fig3}(a) shows the corresponding crystal structure for K$_2$CdPb.
As seen in Table \ref{tab1}, the $\kappa_L$ values predicted by our ML models for the nine compositions A$_2$CdX (A = Li, Na, and K; X = Pb, Sn, and Ge) range from $0.69$ to $0.95$ W/mK, indicating that all these compounds are potential low-$\kappa_L$ materials.

\begin{table*}
\begin{tabular}{cccccccccc}
\hline
\hline
Methods & K$_2$CdPb & K$_2$CdSn & K$_2$CdGe & Na$_2$CdPb & Na$_2$CdSn & Na$_2$CdGe & Li$_2$CdPb & Li$_2$ CdSn & Li$_2$CdGe \\
\hline
ML &0.69&0.79&0.8&0.84&0.76&0.87&0.95&0.71&0.77\\
DFT &0.295&0.346&0.323&0.022&0.119&0.086&0.032&0.215&0.028\\
ML + Weight 5 &0.56&0.35&0.37&0.45&0.63&0.119&0.68&0.27&0.4\\
ML + Weight 10 &0.469&0.29&0.62&0.43&0.37&0.58&0.76&0.4&0.22\\
\hline
\hline
\end{tabular}
\caption{Machine learning (ML) and density functional theory (DFT) predictions of the lattice thermal conductivity $\kappa_L$ (in units of W/mK) for different Cd-compounds. The ML models are based on Random Forests. The terms ``ML + Weight 5" and ``ML + Weight 10" indicate a weighting factor of 5 and 10, respectively, on samples with $\log (\kappa_L) \le 0$ when training the ML models, which places more weight on the low-$\kappa_L$ materials.
\label{tab1}}
\end{table*}

To validate the ML predictions, we further perform first-principles calculations to directly compute $\kappa_L$ and other thermoelectric properties for the nine compounds under study. We first focus on K$_2$CdPb~\cite{zhang2022remarkable}, whose primitive cell structure is shown in Figure \ref{fig3}(a). To ensure accurate calculation of $\kappa_L$, we conduct a convergence test for the neighbor interaction cutoff distance. Figures \ref{fig3}(b) and \ref{fig3}(c) show the convergence tests for K$_2$CdPb as functions of temperature and cutoff distance. Notably, the $\kappa_L$ computed with a cutoff of 4 {\AA} is very close to the result obtained with a 5 {\AA} cutoff. Therefore, for both accuracy and efficiency considerations, we have adopted a cutoff distance of 4 {\AA} for the other compounds. Figure \ref{fig3}(d) shows the computed $\kappa_L$ as a function of temperature for the nine compounds A$_2$CdX (A = Li, Na, and K; X = Pb, Sn, and Ge). Near room temperature, all compounds exhibit a $\kappa_L$ below 1.0 W/mK, in very good agreement with our ML predictions. As the temperature increases, $\kappa_L$ further decreases as more phonons are excited and cause additional phonon scattering, leading to a reduction in $\kappa_L$. These results reveal that {\it the nine compounds under study are all low-$\kappa_L$ materials for potential thermal management and insulation applications}.
We further note that our reported $\kappa_L \sim 0.3$ W/mK for K$_2$CdPb and K$_2$CdSn is consistent with previous DFT studies~\cite{zhang2022remarkable}. Meanwhile, we also find that K$_2$CdGe exhibits a comparable theoretical $\kappa_L \sim 0.3$ W/mK. This result is not surprising, given the chemical similarity of the Pb, Sn, and Ge elements. Likewise, for the Li- and Na-based compounds explored here, while our DFT calculations may underestimate their $\kappa_L$ values, they are also anticipated to be low lattice thermal conductivity materials, based on their chemical similarity to the K-based compounds. In fact, depending on carrier concentration and the underlying temperature, materials with ultra-low $\kappa_L \sim 0.1$ W/mK or lower have been reported in the literature~\cite{koley2021ultralow,fallah2021ultra,gibson2021low,zhang2023extremely}. It would be an important future task to verify our predictions both theoretically (e.g., with different DFT functionals and supercell sizes) and experimentally, through potential synthesis and characterization of the proposed materials.

Before discussing other thermoelectric properties, we address the small discrepancies between the ML and DFT results in Table \ref{tab1}. First, we note that the ML models were trained to predict $\log(\kappa_L)$ rather than $\kappa_L$ itself. Therefore, a small error in the logarithmic value can be amplified in the actual value. 
Second, the ML prediction in Table I is generally slightly larger than the DFT calculation.
One reason for this discrepancy is likely due to the training data distribution. Specifically, while we are interested in discovering materials with low thermal conductivity (i.e., $\log(\kappa_L) \le 0$), most of the training data in Figure \ref{fig1}(a) exhibit $\log(\kappa_L) \ge 0$.
Therefore, one could potentially enhance the ML prediction accuracy by training a weighted ML model. Indeed, when we train additional ML models by weighting the training samples with $\log(\kappa_L) \le 0$ by a factor of 5 to 10, the predicted $\kappa_L$ values become smaller and align more closely with the DFT results, as seen in Table I. Notably, the performance of the ML model with a weight factor of 10 is not better than that with a weight factor of 5. Therefore, the sample weight factor has an optimal range and cannot be increased indefinitely.
Finally, we note that some ML predictions from the weighted ML models remain larger than the DFT results (e.g., for the sodium compounds in Table I). This discrepancy may be attributed to the fact that tree models only interpolate so cannot predict values beyond the range of the training dataset. The apparent, albeit small, differences between the ML and DFT predictions are likely associated with the above factors.

\begin{figure}
\includegraphics[width=1.0\linewidth]{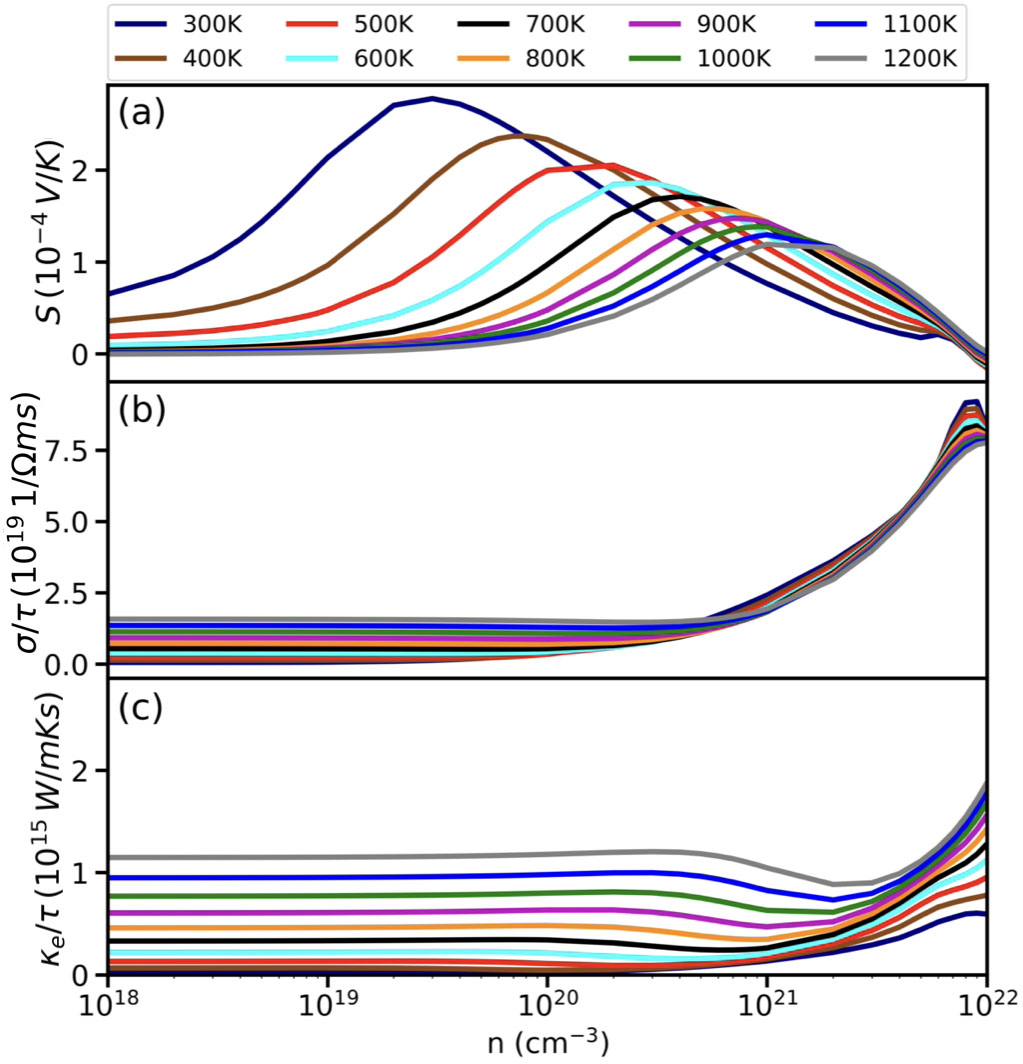}
\caption{Thermoelectric properties of K$_2$CdPb from first-principles calculations:
(\textbf{a}) Seebeck coefficient ($S$), (\textbf{b}) electrical conductivity divided by the relaxation time ($\sigma$/$\tau$), (\textbf{c}) electronic thermal conductivity divided by the relaxation time ($\kappa_{e}$/$\tau$), as a function of the carrier concentration $n$ (in log scale) over the temperature range of $300-1200$ K.\label{fig4}}
\end{figure}

We next turn our attention to the thermoelectric properties. Figure \ref{fig4} displays the DFT calculations for K$_2$CdPb: Seebeck coefficient $S$ [panel (a)], electrical conductivity divided by the relaxation time $\sigma/\tau$ [panel (b)], and electronic thermal conductivity divided by the relaxation time $\kappa_e/\tau$ [panel (c)], as a function of the carrier concentration $n$ in the temperature range $300-1200$ K.
In general, the Seebeck coefficient exhibits a more complex temperature and carrier concentration dependence, but its behavior can be understood qualitatively by considering that of a simple parabolic band~\cite{cutler1964electronic,snyder2008complex}:
\begin{equation}\label{eq:freeS}
S = \frac{8\pi^2 k^2_B}{3eh^2} m^* T (\frac{\pi}{3n})^{2/3}.
\end{equation}
Here, $k_B$, $e$, $h$, and $m^*$ are the Boltzmann constant, electron charge, Plank constant, and carrier effective mass, respectively. 
Equation (\ref{eq:freeS}) dictates that a higher temperature $T$ or a lower carrier concentration $n$ would result in a larger Seebeck coefficient $S$. These $T$ and $n$ dependences are indeed consistent with those shown in Figure \ref{fig4}(a), especially in the high carrier concentration regime ($n > 10^{20}$ cm$^{-3}$). In contrast, the low concentration regime exhibits an opposite trend, where $S$ is reduced at higher temperatures. This anomalous behavior is caused by the bipolar effect~\cite{glassbrenner1964thermal,shi2015connecting,gong2016investigation}, where thermal excitations generate both electrons and holes, which contribute opposite signs and lead to an overall reduced $S$.

The behavior of the electrical conductivity $\sigma$ shown in Figure \ref{fig4}(b) is more straightforward. Specifically, $\sigma$ is anticipated to correlate with $n/m^*$ and show only weak temperature dependence. Additionally, the electronic thermal conductivity $\kappa_{e}$ can be related to $\sigma$ via the Wiedemann-Franz law~\cite{chester1961law}: $\kappa_{e} = L \sigma T$, where $L$ is the Lorentz number ($2.44 \times 10^{-8}$ W$\Omega$/K$^2$ for free electrons). Figure \ref{fig4}(c) shows that $\kappa_{e}$ roughly exhibits a linear relationship with respect to $T$ and $n$, which indeed closely follows the Wiedemann-Franz law.
We note that $\kappa_e$ becomes significantly larger only near $n \sim 10^{22}$ cm$^{-3}$ or at high temperature.
It remains computationally very challenging to directly compute the relaxation time $\tau$ from first principles. Meanwhile, assuming a typical value of $\tau = 1 \times 10^{-14}$ s (also commonly employed in the literature), $\kappa_e$ is less than $1-10$ W/mK in most of the temperature range and carrier concentrations under study. Thus, K$_2$CdPb remains a low-$\kappa$ material even after taking into account the electronic contribution.

\begin{figure}
\includegraphics[width=1.0\linewidth]{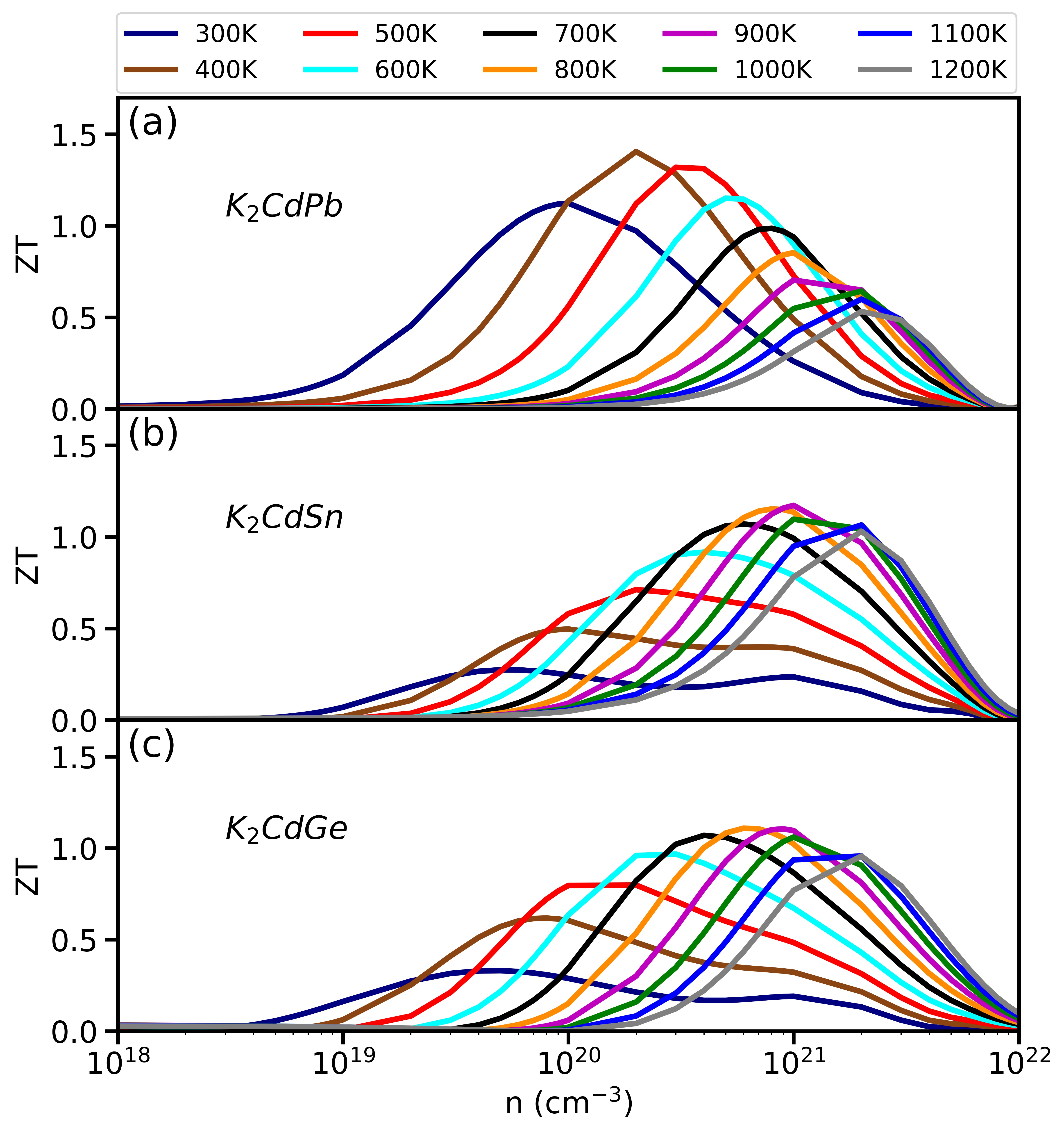}
\caption{$ZT$ values of thermoelectric performance from first-principles calculations for (a) K$_2$CdPb, (b) K$_2$CdSn, and (c) K$_2$CdGe, as a function of the carrier concentration $n$ (in log scale) over the temperature range of $300-1200$ K.
\label{fig5}}
\end{figure}

Finally, since low-$\kappa$ materials can be good candidates for thermoelectrics, we also compute their figure of merit, $ZT = S^2 \sigma T/\kappa$, where the thermal conductivity $\kappa = \kappa_e + \kappa_L$ includes both electronic and lattice contributions.
Figures \ref{fig5}(a)-(c) show the $ZT$ values respectively for K$_2$CdPb, K$_2$CdSn, and K$_2$CdGe, as functions of carrier concentration $n$ (in log scale) and temperature $T$. In all three compounds, the $ZT$ values can exceed 1.0.
As an example to estimate the $ZT$ value, for K$_2$CdPb at $T = 400$ K and $n = 2 \times 10^{20}$ cm$^{-3}$, the relevant parameters from our calculations are $S \sim 1.8 \times 10^{-4}$ V/K, $\sigma \sim 0.1 \times 10^5$ 1/$\Omega$m, and $\kappa = \kappa_e + \kappa_L \sim 1.0$ W/mK. Together, these values lead to a figure of merit $ZT = S^2 \sigma T/\kappa (\sim 3.24 \times10^{-8} \times 0.1 \times 10^5 \times 400 /1.0) \sim 1.3 - 1.4$ for K$_2$CdPb, making it a {\it promising low-temperature thermoelectric material}.
In contrast, K$_2$CdSn and K$_2$CdGe show peak $ZT$ values of $\sim 1.1$ near $n=9 \times 10 ^{20}$ cm$^{-3}$ and  $T=900$ K, and they are more suitable for thermoelectric applications at higher temperatures.
For the other compounds based on sodium and lithium listed in Table I, the $ZT$ values are less than 1.0, making them unsuitable for practical thermoelectric applications. However, they could still be potential candidates for thermal insulation materials.

%\begin{listing}[H]
%\caption{Title of the listing}
%\rule{\columnwidth}{1pt}
%\raggedright Text of the listing. In font size footnotesize, small, or normalsize. Preferred format: left aligned and single spaced. Preferred border format: top border line and bottom border line.
%\rule{\columnwidth}{1pt}
%\end{listing}

%%%%%%%%%%%%%%%%%%%%%%%%%%%%%%%%%%%%%%%%%%
%\section{Discussion}
%Authors should discuss the results and how they can be interpreted from the perspective of previous studies and of the working hypotheses. The findings and their implications should be discussed in the broadest context possible. Future research directions may also be highlighted.

%%%%%%%%%%%%%%%%%%%%%%%%%%%%%%%%%%%%%%%%%%
\section{\label{sec:conclusion}Conclusion}

We have developed machine learning (ML) models using Random Forests to efficiently predict the lattice thermal conductivity ($\kappa_L$) of a given chemical compound. We have also conducted first-principles density functional theory (DFT) calculations to validate the ML predictions. The results indicate that the nine Zintl-phase Cd-compounds A$_2$CdX (A = Li, Na, and K; X = Pb, Sn, and Ge) with orthorhombic crystal symmetry all exhibit very low lattice thermal conductivities, with $\kappa_L \leq 1.0$ W/mK. Our DFT calculations of the figure of merit, $ZT$, for thermoelectric performance further showed that K$_2$CdPb exhibits a peak $ZT \sim 1.4$ near 400 Kelvin, making it a promising low-temperature thermoelectric material. Additionally, K$_2$CdSn and K$_2$CdGe were found to display $ZT$ values of  $\sim 1.1$ at 900 Kelvin, suggesting they could be candidate thermoelectrics at higher temperatures.

For Li$_2$CdX and Na$_2$CdX (X = Pb, Sn, and Ge), the $ZT$ values are less than 1.0, indicating more limited practical thermoelectric applications. Nevertheless, their ultra-low lattice thermal conductivities make these materials potentially useful for thermal management and insulation applications. Overall, our study demonstrated that data-driven ML methods are powerful tools for large-scale materials modeling and discovery. Experimental verification of our ML and DFT predictions on the thermoelectric properties of the Zintl-phase Cd-compounds would be an important next step. Further theoretical exploration of additional low-$\kappa_L$ and high-$ZT$ materials using a combined ML and DFT methodology will continue to be an important area of future research.

\section*{Acknowledgements}
This work is supported by the U.S. Air Force Office of Scientific Research (AFOSR) under Award No.
FA2386-21-1-4060.
The calculations were performed on the Frontera supercomputer at the Texas Advanced Computing Center. Frontera is made possible by National Science Foundation Award No. OAC-1818253.

\section*{Data Availability Statement}
The data for training machine learning (ML) models and the resulting ML code for predicting lattice thermal conductivity can be found online at the following weblink:
{\url{https://github.com/CMLUAB/ML\_lattice-themal-conductivity}}.
The data for first-principles calculations are available upon request from the authors.

%\section*{Competing Financial Statement}
%The authors declare no conflict of financial interests.

\section*{References}

\bibliography{bibfile}

\begin{thebibliography}{89}%
\makeatletter
\providecommand \@ifxundefined [1]{%
 \@ifx{#1\undefined}
}%
\providecommand \@ifnum [1]{%
 \ifnum #1\expandafter \@firstoftwo
 \else \expandafter \@secondoftwo
 \fi
}%
\providecommand \@ifx [1]{%
 \ifx #1\expandafter \@firstoftwo
 \else \expandafter \@secondoftwo
 \fi
}%
\providecommand \natexlab [1]{#1}%
\providecommand \enquote  [1]{``#1''}%
\providecommand \bibnamefont  [1]{#1}%
\providecommand \bibfnamefont [1]{#1}%
\providecommand \citenamefont [1]{#1}%
\providecommand \href@noop [0]{\@secondoftwo}%
\providecommand \href [0]{\begingroup \@sanitize@url \@href}%
\providecommand \@href[1]{\@@startlink{#1}\@@href}%
\providecommand \@@href[1]{\endgroup#1\@@endlink}%
\providecommand \@sanitize@url [0]{\catcode `\\12\catcode `\$12\catcode
  `\&12\catcode `\#12\catcode `\^12\catcode `\_12\catcode `\%12\relax}%
\providecommand \@@startlink[1]{}%
\providecommand \@@endlink[0]{}%
\providecommand \url  [0]{\begingroup\@sanitize@url \@url }%
\providecommand \@url [1]{\endgroup\@href {#1}{\urlprefix }}%
\providecommand \urlprefix  [0]{URL }%
\providecommand \Eprint [0]{\href }%
\providecommand \doibase [0]{https://doi.org/}%
\providecommand \selectlanguage [0]{\@gobble}%
\providecommand \bibinfo  [0]{\@secondoftwo}%
\providecommand \bibfield  [0]{\@secondoftwo}%
\providecommand \translation [1]{[#1]}%
\providecommand \BibitemOpen [0]{}%
\providecommand \bibitemStop [0]{}%
\providecommand \bibitemNoStop [0]{.\EOS\space}%
\providecommand \EOS [0]{\spacefactor3000\relax}%
\providecommand \BibitemShut  [1]{\csname bibitem#1\endcsname}%
\let\auto@bib@innerbib\@empty
%</preamble>
\bibitem [{\citenamefont {He}\ and\ \citenamefont
  {Tritt}(2017)}]{he2017advances}%
  \BibitemOpen
  \bibfield  {author} {\bibinfo {author} {\bibfnamefont {J.}~\bibnamefont
  {He}}\ and\ \bibinfo {author} {\bibfnamefont {T.~M.}\ \bibnamefont {Tritt}},\
  }\bibfield  {title} {\bibinfo {title} {Advances in thermoelectric materials
  research: Looking back and moving forward},\ }\href
  {https://doi.org/10.1126/science.aak9997} {\bibfield  {journal} {\bibinfo
  {journal} {Science}\ }\textbf {\bibinfo {volume} {357}},\ \bibinfo {pages}
  {eaak9997} (\bibinfo {year} {2017})}\BibitemShut {NoStop}%
\bibitem [{\citenamefont {Liu}\ \emph {et~al.}(2017)\citenamefont {Liu},
  \citenamefont {Hu}, \citenamefont {Zhang}, \citenamefont {Deng},
  \citenamefont {Han},\ and\ \citenamefont {Liu}}]{liu2017new}%
  \BibitemOpen
  \bibfield  {author} {\bibinfo {author} {\bibfnamefont {W.}~\bibnamefont
  {Liu}}, \bibinfo {author} {\bibfnamefont {J.}~\bibnamefont {Hu}}, \bibinfo
  {author} {\bibfnamefont {S.}~\bibnamefont {Zhang}}, \bibinfo {author}
  {\bibfnamefont {M.}~\bibnamefont {Deng}}, \bibinfo {author} {\bibfnamefont
  {C.-G.}\ \bibnamefont {Han}},\ and\ \bibinfo {author} {\bibfnamefont
  {Y.}~\bibnamefont {Liu}},\ }\bibfield  {title} {\bibinfo {title} {New trends,
  strategies and opportunities in thermoelectric materials: a perspective},\
  }\href {https://doi.org/10.1016/j.mtphys.2017.06.001} {\bibfield  {journal}
  {\bibinfo  {journal} {Materials Today Physics}\ }\textbf {\bibinfo {volume}
  {1}},\ \bibinfo {pages} {50} (\bibinfo {year} {2017})}\BibitemShut {NoStop}%
\bibitem [{\citenamefont {Zevalkink}\ \emph {et~al.}(2018)\citenamefont
  {Zevalkink}, \citenamefont {Smiadak}, \citenamefont {Blackburn},
  \citenamefont {Ferguson}, \citenamefont {Chabinyc}, \citenamefont {Delaire},
  \citenamefont {Wang}, \citenamefont {Kovnir}, \citenamefont {Martin},
  \citenamefont {Schelhas} \emph {et~al.}}]{zevalkink2018practical}%
  \BibitemOpen
  \bibfield  {author} {\bibinfo {author} {\bibfnamefont {A.}~\bibnamefont
  {Zevalkink}}, \bibinfo {author} {\bibfnamefont {D.~M.}\ \bibnamefont
  {Smiadak}}, \bibinfo {author} {\bibfnamefont {J.~L.}\ \bibnamefont
  {Blackburn}}, \bibinfo {author} {\bibfnamefont {A.~J.}\ \bibnamefont
  {Ferguson}}, \bibinfo {author} {\bibfnamefont {M.~L.}\ \bibnamefont
  {Chabinyc}}, \bibinfo {author} {\bibfnamefont {O.}~\bibnamefont {Delaire}},
  \bibinfo {author} {\bibfnamefont {J.}~\bibnamefont {Wang}}, \bibinfo {author}
  {\bibfnamefont {K.}~\bibnamefont {Kovnir}}, \bibinfo {author} {\bibfnamefont
  {J.}~\bibnamefont {Martin}}, \bibinfo {author} {\bibfnamefont {L.~T.}\
  \bibnamefont {Schelhas}}, \emph {et~al.},\ }\bibfield  {title} {\bibinfo
  {title} {A practical field guide to thermoelectrics: Fundamentals, synthesis,
  and characterization},\ }\bibfield  {journal} {\bibinfo  {journal} {Applied
  Physics Reviews}\ }\textbf {\bibinfo {volume} {5}},\ \href
  {https://doi.org/10.1063/1.5021094} {10.1063/1.5021094} (\bibinfo {year}
  {2018})\BibitemShut {NoStop}%
\bibitem [{\citenamefont {Urban}\ \emph {et~al.}(2019)\citenamefont {Urban},
  \citenamefont {Menon}, \citenamefont {Tian}, \citenamefont {Jain},\ and\
  \citenamefont {Hippalgaonkar}}]{urban2019new}%
  \BibitemOpen
  \bibfield  {author} {\bibinfo {author} {\bibfnamefont {J.~J.}\ \bibnamefont
  {Urban}}, \bibinfo {author} {\bibfnamefont {A.~K.}\ \bibnamefont {Menon}},
  \bibinfo {author} {\bibfnamefont {Z.}~\bibnamefont {Tian}}, \bibinfo {author}
  {\bibfnamefont {A.}~\bibnamefont {Jain}},\ and\ \bibinfo {author}
  {\bibfnamefont {K.}~\bibnamefont {Hippalgaonkar}},\ }\bibfield  {title}
  {\bibinfo {title} {New horizons in thermoelectric materials: Correlated
  electrons, organic transport, machine learning, and more},\ }\bibfield
  {journal} {\bibinfo  {journal} {Journal of Applied Physics}\ }\textbf
  {\bibinfo {volume} {125}},\ \href {https://doi.org/10.1063/1.5092525}
  {10.1063/1.5092525} (\bibinfo {year} {2019})\BibitemShut {NoStop}%
\bibitem [{\citenamefont {Wei}\ \emph {et~al.}(2020)\citenamefont {Wei},
  \citenamefont {Yang}, \citenamefont {Ma}, \citenamefont {Song}, \citenamefont
  {Zhang}, \citenamefont {Ma}, \citenamefont {Yang},\ and\ \citenamefont
  {Wang}}]{wei2020review}%
  \BibitemOpen
  \bibfield  {author} {\bibinfo {author} {\bibfnamefont {J.}~\bibnamefont
  {Wei}}, \bibinfo {author} {\bibfnamefont {L.}~\bibnamefont {Yang}}, \bibinfo
  {author} {\bibfnamefont {Z.}~\bibnamefont {Ma}}, \bibinfo {author}
  {\bibfnamefont {P.}~\bibnamefont {Song}}, \bibinfo {author} {\bibfnamefont
  {M.}~\bibnamefont {Zhang}}, \bibinfo {author} {\bibfnamefont
  {J.}~\bibnamefont {Ma}}, \bibinfo {author} {\bibfnamefont {F.}~\bibnamefont
  {Yang}},\ and\ \bibinfo {author} {\bibfnamefont {X.}~\bibnamefont {Wang}},\
  }\bibfield  {title} {\bibinfo {title} {Review of current high-{ZT}
  thermoelectric materials},\ }\href
  {https://doi.org/10.1007/s10853-020-04949-0} {\bibfield  {journal} {\bibinfo
  {journal} {Journal of Materials Science}\ }\textbf {\bibinfo {volume} {55}},\
  \bibinfo {pages} {12642} (\bibinfo {year} {2020})}\BibitemShut {NoStop}%
\bibitem [{\citenamefont {Hasan}\ \emph {et~al.}(2020)\citenamefont {Hasan},
  \citenamefont {Wahid}, \citenamefont {Nayan},\ and\ \citenamefont
  {Mohamed~Ali}}]{hasan2020inorganic}%
  \BibitemOpen
  \bibfield  {author} {\bibinfo {author} {\bibfnamefont {M.~N.}\ \bibnamefont
  {Hasan}}, \bibinfo {author} {\bibfnamefont {H.}~\bibnamefont {Wahid}},
  \bibinfo {author} {\bibfnamefont {N.}~\bibnamefont {Nayan}},\ and\ \bibinfo
  {author} {\bibfnamefont {M.~S.}\ \bibnamefont {Mohamed~Ali}},\ }\bibfield
  {title} {\bibinfo {title} {Inorganic thermoelectric materials: A review},\
  }\href {https://doi.org/10.1002/er.5313} {\bibfield  {journal} {\bibinfo
  {journal} {International Journal of Energy Research}\ }\textbf {\bibinfo
  {volume} {44}},\ \bibinfo {pages} {6170} (\bibinfo {year}
  {2020})}\BibitemShut {NoStop}%
\bibitem [{\citenamefont {Zoui}\ \emph {et~al.}(2020)\citenamefont {Zoui},
  \citenamefont {Bentouba}, \citenamefont {Stocholm},\ and\ \citenamefont
  {Bourouis}}]{zoui2020review}%
  \BibitemOpen
  \bibfield  {author} {\bibinfo {author} {\bibfnamefont {M.~A.}\ \bibnamefont
  {Zoui}}, \bibinfo {author} {\bibfnamefont {S.}~\bibnamefont {Bentouba}},
  \bibinfo {author} {\bibfnamefont {J.~G.}\ \bibnamefont {Stocholm}},\ and\
  \bibinfo {author} {\bibfnamefont {M.}~\bibnamefont {Bourouis}},\ }\bibfield
  {title} {\bibinfo {title} {A review on thermoelectric generators: Progress
  and applications},\ }\href {https://doi.org/10.3390/en13143606} {\bibfield
  {journal} {\bibinfo  {journal} {Energies}\ }\textbf {\bibinfo {volume}
  {13}},\ \bibinfo {pages} {3606} (\bibinfo {year} {2020})}\BibitemShut
  {NoStop}%
\bibitem [{\citenamefont {Jaziri}\ \emph {et~al.}(2020)\citenamefont {Jaziri},
  \citenamefont {Boughamoura}, \citenamefont {M{\"u}ller}, \citenamefont
  {Mezghani}, \citenamefont {Tounsi},\ and\ \citenamefont
  {Ismail}}]{jaziri2020comprehensive}%
  \BibitemOpen
  \bibfield  {author} {\bibinfo {author} {\bibfnamefont {N.}~\bibnamefont
  {Jaziri}}, \bibinfo {author} {\bibfnamefont {A.}~\bibnamefont {Boughamoura}},
  \bibinfo {author} {\bibfnamefont {J.}~\bibnamefont {M{\"u}ller}}, \bibinfo
  {author} {\bibfnamefont {B.}~\bibnamefont {Mezghani}}, \bibinfo {author}
  {\bibfnamefont {F.}~\bibnamefont {Tounsi}},\ and\ \bibinfo {author}
  {\bibfnamefont {M.}~\bibnamefont {Ismail}},\ }\bibfield  {title} {\bibinfo
  {title} {A comprehensive review of thermoelectric generators: Technologies
  and common applications},\ }\href
  {https://doi.org/10.1016/j.egyr.2019.12.011} {\bibfield  {journal} {\bibinfo
  {journal} {Energy Reports}\ }\textbf {\bibinfo {volume} {6}},\ \bibinfo
  {pages} {264} (\bibinfo {year} {2020})}\BibitemShut {NoStop}%
\bibitem [{\citenamefont {Pei}\ \emph {et~al.}(2014)\citenamefont {Pei},
  \citenamefont {Wu}, \citenamefont {Wu}, \citenamefont {Zheng},\ and\
  \citenamefont {He}}]{pei2014high}%
  \BibitemOpen
  \bibfield  {author} {\bibinfo {author} {\bibfnamefont {Y.-L.}\ \bibnamefont
  {Pei}}, \bibinfo {author} {\bibfnamefont {H.}~\bibnamefont {Wu}}, \bibinfo
  {author} {\bibfnamefont {D.}~\bibnamefont {Wu}}, \bibinfo {author}
  {\bibfnamefont {F.}~\bibnamefont {Zheng}},\ and\ \bibinfo {author}
  {\bibfnamefont {J.}~\bibnamefont {He}},\ }\bibfield  {title} {\bibinfo
  {title} {{High thermoelectric performance realized in a BiCuSeO system by
  improving carrier mobility through 3D modulation doping}},\ }\href
  {https://doi.org/10.1021/ja507945h} {\bibfield  {journal} {\bibinfo
  {journal} {Journal of the American Chemical Society}\ }\textbf {\bibinfo
  {volume} {136}},\ \bibinfo {pages} {13902} (\bibinfo {year}
  {2014})}\BibitemShut {NoStop}%
\bibitem [{\citenamefont {Lee}\ \emph {et~al.}(2015)\citenamefont {Lee},
  \citenamefont {Kim}, \citenamefont {Mun}, \citenamefont {Ryu}, \citenamefont
  {Choi}, \citenamefont {Park}, \citenamefont {Hwang},\ and\ \citenamefont
  {Kim}}]{lee2015enhanced}%
  \BibitemOpen
  \bibfield  {author} {\bibinfo {author} {\bibfnamefont {K.~H.}\ \bibnamefont
  {Lee}}, \bibinfo {author} {\bibfnamefont {S.~I.}\ \bibnamefont {Kim}},
  \bibinfo {author} {\bibfnamefont {H.}~\bibnamefont {Mun}}, \bibinfo {author}
  {\bibfnamefont {B.}~\bibnamefont {Ryu}}, \bibinfo {author} {\bibfnamefont
  {S.-M.}\ \bibnamefont {Choi}}, \bibinfo {author} {\bibfnamefont {H.~J.}\
  \bibnamefont {Park}}, \bibinfo {author} {\bibfnamefont {S.}~\bibnamefont
  {Hwang}},\ and\ \bibinfo {author} {\bibfnamefont {S.~W.}\ \bibnamefont
  {Kim}},\ }\bibfield  {title} {\bibinfo {title} {{Enhanced thermoelectric
  performance of n-type Cu0.008Bi2Te2.7Se0.3 by band engineering}},\ }\href
  {https://doi.org/10.1021/ja507945h} {\bibfield  {journal} {\bibinfo
  {journal} {Journal of Materials Chemistry C}\ }\textbf {\bibinfo {volume}
  {3}},\ \bibinfo {pages} {10604} (\bibinfo {year} {2015})}\BibitemShut
  {NoStop}%
\bibitem [{\citenamefont {Lu}\ \emph {et~al.}(2015)\citenamefont {Lu},
  \citenamefont {Morelli}, \citenamefont {Xia},\ and\ \citenamefont
  {Ozolins}}]{lu2015increasing}%
  \BibitemOpen
  \bibfield  {author} {\bibinfo {author} {\bibfnamefont {X.}~\bibnamefont
  {Lu}}, \bibinfo {author} {\bibfnamefont {D.~T.}\ \bibnamefont {Morelli}},
  \bibinfo {author} {\bibfnamefont {Y.}~\bibnamefont {Xia}},\ and\ \bibinfo
  {author} {\bibfnamefont {V.}~\bibnamefont {Ozolins}},\ }\bibfield  {title}
  {\bibinfo {title} {Increasing the thermoelectric figure of merit of
  tetrahedrites by co-doping with nickel and zinc},\ }\href
  {https://doi.org/10.1021/cm502570b} {\bibfield  {journal} {\bibinfo
  {journal} {Chemistry of Materials}\ }\textbf {\bibinfo {volume} {27}},\
  \bibinfo {pages} {408} (\bibinfo {year} {2015})}\BibitemShut {NoStop}%
\bibitem [{\citenamefont {Jiang}\ \emph {et~al.}(2021)\citenamefont {Jiang},
  \citenamefont {Yu}, \citenamefont {Chen}, \citenamefont {Cui}, \citenamefont
  {Liu}, \citenamefont {Xie},\ and\ \citenamefont {He}}]{jiang2021entropy}%
  \BibitemOpen
  \bibfield  {author} {\bibinfo {author} {\bibfnamefont {B.}~\bibnamefont
  {Jiang}}, \bibinfo {author} {\bibfnamefont {Y.}~\bibnamefont {Yu}}, \bibinfo
  {author} {\bibfnamefont {H.}~\bibnamefont {Chen}}, \bibinfo {author}
  {\bibfnamefont {J.}~\bibnamefont {Cui}}, \bibinfo {author} {\bibfnamefont
  {X.}~\bibnamefont {Liu}}, \bibinfo {author} {\bibfnamefont {L.}~\bibnamefont
  {Xie}},\ and\ \bibinfo {author} {\bibfnamefont {J.}~\bibnamefont {He}},\
  }\bibfield  {title} {\bibinfo {title} {Entropy engineering promotes
  thermoelectric performance in p-type chalcogenides},\ }\href
  {https://doi.org/10.1038/s41467-021-23569-z} {\bibfield  {journal} {\bibinfo
  {journal} {Nature Communications}\ }\textbf {\bibinfo {volume} {12}},\
  \bibinfo {pages} {3234} (\bibinfo {year} {2021})}\BibitemShut {NoStop}%
\bibitem [{\citenamefont {Ma}\ \emph {et~al.}(2021)\citenamefont {Ma},
  \citenamefont {Wei}, \citenamefont {Song}, \citenamefont {Zhang},
  \citenamefont {Yang}, \citenamefont {Ma}, \citenamefont {Liu}, \citenamefont
  {Yang},\ and\ \citenamefont {Wang}}]{ma2021review}%
  \BibitemOpen
  \bibfield  {author} {\bibinfo {author} {\bibfnamefont {Z.}~\bibnamefont
  {Ma}}, \bibinfo {author} {\bibfnamefont {J.}~\bibnamefont {Wei}}, \bibinfo
  {author} {\bibfnamefont {P.}~\bibnamefont {Song}}, \bibinfo {author}
  {\bibfnamefont {M.}~\bibnamefont {Zhang}}, \bibinfo {author} {\bibfnamefont
  {L.}~\bibnamefont {Yang}}, \bibinfo {author} {\bibfnamefont {J.}~\bibnamefont
  {Ma}}, \bibinfo {author} {\bibfnamefont {W.}~\bibnamefont {Liu}}, \bibinfo
  {author} {\bibfnamefont {F.}~\bibnamefont {Yang}},\ and\ \bibinfo {author}
  {\bibfnamefont {X.}~\bibnamefont {Wang}},\ }\bibfield  {title} {\bibinfo
  {title} {Review of experimental approaches for improving zt of thermoelectric
  materials},\ }\href {https://doi.org/10.1016/j.mssp.2020.105303} {\bibfield
  {journal} {\bibinfo  {journal} {Materials Science in Semiconductor
  Processing}\ }\textbf {\bibinfo {volume} {121}},\ \bibinfo {pages} {105303}
  (\bibinfo {year} {2021})}\BibitemShut {NoStop}%
\bibitem [{\citenamefont {Ghosh}\ \emph {et~al.}(2022)\citenamefont {Ghosh},
  \citenamefont {Dutta}, \citenamefont {Sarkar},\ and\ \citenamefont
  {Biswas}}]{ghosh2022insights}%
  \BibitemOpen
  \bibfield  {author} {\bibinfo {author} {\bibfnamefont {T.}~\bibnamefont
  {Ghosh}}, \bibinfo {author} {\bibfnamefont {M.}~\bibnamefont {Dutta}},
  \bibinfo {author} {\bibfnamefont {D.}~\bibnamefont {Sarkar}},\ and\ \bibinfo
  {author} {\bibfnamefont {K.}~\bibnamefont {Biswas}},\ }\bibfield  {title}
  {\bibinfo {title} {Insights into low thermal conductivity in inorganic
  materials for thermoelectrics},\ }\href
  {https://doi.org/10.1021/jacs.2c02017} {\bibfield  {journal} {\bibinfo
  {journal} {Journal of the American Chemical Society}\ }\textbf {\bibinfo
  {volume} {144}},\ \bibinfo {pages} {10099} (\bibinfo {year}
  {2022})}\BibitemShut {NoStop}%
\bibitem [{\citenamefont {Ding}\ \emph {et~al.}(2023)\citenamefont {Ding},
  \citenamefont {Duan}, \citenamefont {Ding}, \citenamefont {Pan},
  \citenamefont {Wang}, \citenamefont {Xiao}, \citenamefont {Liu},
  \citenamefont {Li}, \citenamefont {Luo}, \citenamefont {Zeng} \emph
  {et~al.}}]{ding2023xmosin2}%
  \BibitemOpen
  \bibfield  {author} {\bibinfo {author} {\bibfnamefont {C.-H.}\ \bibnamefont
  {Ding}}, \bibinfo {author} {\bibfnamefont {Z.-F.}\ \bibnamefont {Duan}},
  \bibinfo {author} {\bibfnamefont {Z.-K.}\ \bibnamefont {Ding}}, \bibinfo
  {author} {\bibfnamefont {H.}~\bibnamefont {Pan}}, \bibinfo {author}
  {\bibfnamefont {J.}~\bibnamefont {Wang}}, \bibinfo {author} {\bibfnamefont
  {W.-H.}\ \bibnamefont {Xiao}}, \bibinfo {author} {\bibfnamefont {W.-P.}\
  \bibnamefont {Liu}}, \bibinfo {author} {\bibfnamefont {Q.-Q.}\ \bibnamefont
  {Li}}, \bibinfo {author} {\bibfnamefont {N.-N.}\ \bibnamefont {Luo}},
  \bibinfo {author} {\bibfnamefont {J.}~\bibnamefont {Zeng}}, \emph {et~al.},\
  }\bibfield  {title} {\bibinfo {title} {{XMoSiN2 (X= S, Se, Te): A novel 2D
  Janus semiconductor with ultra-high carrier mobility and excellent
  thermoelectric performance}},\ }\href
  {https://doi.org/10.1209/0295-5075/acdb98} {\bibfield  {journal} {\bibinfo
  {journal} {Europhysics Letters}\ }\textbf {\bibinfo {volume} {143}},\
  \bibinfo {pages} {16002} (\bibinfo {year} {2023})}\BibitemShut {NoStop}%
\bibitem [{\citenamefont {Zhu}\ \emph {et~al.}(2021)\citenamefont {Zhu},
  \citenamefont {He}, \citenamefont {Gong}, \citenamefont {Xie}, \citenamefont
  {Gorai}, \citenamefont {Nielsch},\ and\ \citenamefont
  {Grossman}}]{zhu2021charting}%
  \BibitemOpen
  \bibfield  {author} {\bibinfo {author} {\bibfnamefont {T.}~\bibnamefont
  {Zhu}}, \bibinfo {author} {\bibfnamefont {R.}~\bibnamefont {He}}, \bibinfo
  {author} {\bibfnamefont {S.}~\bibnamefont {Gong}}, \bibinfo {author}
  {\bibfnamefont {T.}~\bibnamefont {Xie}}, \bibinfo {author} {\bibfnamefont
  {P.}~\bibnamefont {Gorai}}, \bibinfo {author} {\bibfnamefont
  {K.}~\bibnamefont {Nielsch}},\ and\ \bibinfo {author} {\bibfnamefont {J.~C.}\
  \bibnamefont {Grossman}},\ }\bibfield  {title} {\bibinfo {title} {Charting
  lattice thermal conductivity for inorganic crystals and discovering rare
  earth chalcogenides for thermoelectrics},\ }\href
  {https://doi.org/10.1039/D1EE00442E} {\bibfield  {journal} {\bibinfo
  {journal} {Energy \& Environmental Science}\ }\textbf {\bibinfo {volume}
  {14}},\ \bibinfo {pages} {3559} (\bibinfo {year} {2021})}\BibitemShut
  {NoStop}%
\bibitem [{\citenamefont {Loftis}\ \emph {et~al.}(2020)\citenamefont {Loftis},
  \citenamefont {Yuan}, \citenamefont {Zhao}, \citenamefont {Hu},\ and\
  \citenamefont {Hu}}]{loftis2020lattice}%
  \BibitemOpen
  \bibfield  {author} {\bibinfo {author} {\bibfnamefont {C.}~\bibnamefont
  {Loftis}}, \bibinfo {author} {\bibfnamefont {K.}~\bibnamefont {Yuan}},
  \bibinfo {author} {\bibfnamefont {Y.}~\bibnamefont {Zhao}}, \bibinfo {author}
  {\bibfnamefont {M.}~\bibnamefont {Hu}},\ and\ \bibinfo {author}
  {\bibfnamefont {J.}~\bibnamefont {Hu}},\ }\bibfield  {title} {\bibinfo
  {title} {Lattice thermal conductivity prediction using symbolic regression
  and machine learning},\ }\href {https://doi.org/10.1021/acs.jpca.0c08103}
  {\bibfield  {journal} {\bibinfo  {journal} {The Journal of Physical Chemistry
  A}\ }\textbf {\bibinfo {volume} {125}},\ \bibinfo {pages} {435} (\bibinfo
  {year} {2020})}\BibitemShut {NoStop}%
\bibitem [{\citenamefont {Tran{\aa}s}\ \emph {et~al.}(2022)\citenamefont
  {Tran{\aa}s}, \citenamefont {L{\o}vvik}, \citenamefont {Tomic},\ and\
  \citenamefont {Berland}}]{tranaas2022lattice}%
  \BibitemOpen
  \bibfield  {author} {\bibinfo {author} {\bibfnamefont {R.}~\bibnamefont
  {Tran{\aa}s}}, \bibinfo {author} {\bibfnamefont {O.~M.}\ \bibnamefont
  {L{\o}vvik}}, \bibinfo {author} {\bibfnamefont {O.}~\bibnamefont {Tomic}},\
  and\ \bibinfo {author} {\bibfnamefont {K.}~\bibnamefont {Berland}},\
  }\bibfield  {title} {\bibinfo {title} {Lattice thermal conductivity of
  half-heuslers with density functional theory and machine learning: Enhancing
  predictivity by active sampling with principal component analysis},\ }\href
  {https://doi.org/10.1016/j.commatsci.2021.110938} {\bibfield  {journal}
  {\bibinfo  {journal} {Computational Materials Science}\ }\textbf {\bibinfo
  {volume} {202}},\ \bibinfo {pages} {110938} (\bibinfo {year}
  {2022})}\BibitemShut {NoStop}%
\bibitem [{\citenamefont {Chester}\ and\ \citenamefont
  {Thellung}(1961)}]{chester1961law}%
  \BibitemOpen
  \bibfield  {author} {\bibinfo {author} {\bibfnamefont {G.}~\bibnamefont
  {Chester}}\ and\ \bibinfo {author} {\bibfnamefont {A.}~\bibnamefont
  {Thellung}},\ }\bibfield  {title} {\bibinfo {title} {The law of wiedemann and
  franz},\ }\href {https://doi.org/10.1088/0370-1328/77/5/309} {\bibfield
  {journal} {\bibinfo  {journal} {Proceedings of the Physical Society}\
  }\textbf {\bibinfo {volume} {77}},\ \bibinfo {pages} {1005} (\bibinfo {year}
  {1961})}\BibitemShut {NoStop}%
\bibitem [{\citenamefont {Seko}\ \emph {et~al.}(2015)\citenamefont {Seko},
  \citenamefont {Togo}, \citenamefont {Hayashi}, \citenamefont {Tsuda},
  \citenamefont {Chaput},\ and\ \citenamefont {Tanaka}}]{seko2015prediction}%
  \BibitemOpen
  \bibfield  {author} {\bibinfo {author} {\bibfnamefont {A.}~\bibnamefont
  {Seko}}, \bibinfo {author} {\bibfnamefont {A.}~\bibnamefont {Togo}}, \bibinfo
  {author} {\bibfnamefont {H.}~\bibnamefont {Hayashi}}, \bibinfo {author}
  {\bibfnamefont {K.}~\bibnamefont {Tsuda}}, \bibinfo {author} {\bibfnamefont
  {L.}~\bibnamefont {Chaput}},\ and\ \bibinfo {author} {\bibfnamefont
  {I.}~\bibnamefont {Tanaka}},\ }\bibfield  {title} {\bibinfo {title}
  {Prediction of low-thermal-conductivity compounds with first-principles
  anharmonic lattice-dynamics calculations and bayesian optimization},\ }\href
  {https://doi.org/10.1103/PhysRevLett.115.205901} {\bibfield  {journal}
  {\bibinfo  {journal} {Physical Review Letters}\ }\textbf {\bibinfo {volume}
  {115}},\ \bibinfo {pages} {205901} (\bibinfo {year} {2015})}\BibitemShut
  {NoStop}%
\bibitem [{\citenamefont {Yang}\ \emph {et~al.}(2021)\citenamefont {Yang},
  \citenamefont {Huh}, \citenamefont {Ning}, \citenamefont {Rapp},
  \citenamefont {Zeng}, \citenamefont {Liu}, \citenamefont {Ju}, \citenamefont
  {Tao}, \citenamefont {Jiang}, \citenamefont {Beak} \emph
  {et~al.}}]{yang2021high}%
  \BibitemOpen
  \bibfield  {author} {\bibinfo {author} {\bibfnamefont {L.}~\bibnamefont
  {Yang}}, \bibinfo {author} {\bibfnamefont {D.}~\bibnamefont {Huh}}, \bibinfo
  {author} {\bibfnamefont {R.}~\bibnamefont {Ning}}, \bibinfo {author}
  {\bibfnamefont {V.}~\bibnamefont {Rapp}}, \bibinfo {author} {\bibfnamefont
  {Y.}~\bibnamefont {Zeng}}, \bibinfo {author} {\bibfnamefont {Y.}~\bibnamefont
  {Liu}}, \bibinfo {author} {\bibfnamefont {S.}~\bibnamefont {Ju}}, \bibinfo
  {author} {\bibfnamefont {Y.}~\bibnamefont {Tao}}, \bibinfo {author}
  {\bibfnamefont {Y.}~\bibnamefont {Jiang}}, \bibinfo {author} {\bibfnamefont
  {J.}~\bibnamefont {Beak}}, \emph {et~al.},\ }\bibfield  {title} {\bibinfo
  {title} {High thermoelectric figure of merit of porous si nanowires from 300
  to 700 k},\ }\href {https://doi.org/10.1038/s41467-021-24208-3} {\bibfield
  {journal} {\bibinfo  {journal} {Nature Communications}\ }\textbf {\bibinfo
  {volume} {12}},\ \bibinfo {pages} {3926} (\bibinfo {year}
  {2021})}\BibitemShut {NoStop}%
\bibitem [{\citenamefont {Lin}\ \emph {et~al.}(2021)\citenamefont {Lin},
  \citenamefont {Chen},\ and\ \citenamefont {Chen}}]{lin2021first}%
  \BibitemOpen
  \bibfield  {author} {\bibinfo {author} {\bibfnamefont {C.-M.}\ \bibnamefont
  {Lin}}, \bibinfo {author} {\bibfnamefont {W.-C.}\ \bibnamefont {Chen}},\ and\
  \bibinfo {author} {\bibfnamefont {C.-C.}\ \bibnamefont {Chen}},\ }\bibfield
  {title} {\bibinfo {title} {{First-principles study of strain effect on the
  thermoelectric properties of LaP and LaAs}},\ }\href
  {https://doi.org/10.1039/D1CP02871E} {\bibfield  {journal} {\bibinfo
  {journal} {Physical Chemistry Chemical Physics}\ }\textbf {\bibinfo {volume}
  {23}},\ \bibinfo {pages} {18189} (\bibinfo {year} {2021})}\BibitemShut
  {NoStop}%
\bibitem [{\citenamefont {Wu}\ \emph {et~al.}(2022)\citenamefont {Wu},
  \citenamefont {Ren}, \citenamefont {Xie}, \citenamefont {Zhou}, \citenamefont
  {Zhang},\ and\ \citenamefont {Chen}}]{wu2022enhanced}%
  \BibitemOpen
  \bibfield  {author} {\bibinfo {author} {\bibfnamefont {C.-W.}\ \bibnamefont
  {Wu}}, \bibinfo {author} {\bibfnamefont {X.}~\bibnamefont {Ren}}, \bibinfo
  {author} {\bibfnamefont {G.}~\bibnamefont {Xie}}, \bibinfo {author}
  {\bibfnamefont {W.-X.}\ \bibnamefont {Zhou}}, \bibinfo {author}
  {\bibfnamefont {G.}~\bibnamefont {Zhang}},\ and\ \bibinfo {author}
  {\bibfnamefont {K.-Q.}\ \bibnamefont {Chen}},\ }\bibfield  {title} {\bibinfo
  {title} {Enhanced high-temperature thermoelectric performance by strain
  engineering in biocl},\ }\href
  {https://doi.org/10.1103/PhysRevApplied.18.014053} {\bibfield  {journal}
  {\bibinfo  {journal} {Physical Review Applied}\ }\textbf {\bibinfo {volume}
  {18}},\ \bibinfo {pages} {014053} (\bibinfo {year} {2022})}\BibitemShut
  {NoStop}%
\bibitem [{\citenamefont {Govindaraj}\ \emph {et~al.}(2022)\citenamefont
  {Govindaraj}, \citenamefont {Sivasamy}, \citenamefont {Murugan},
  \citenamefont {Venugopal},\ and\ \citenamefont
  {Veluswamy}}]{govindaraj2022pressure}%
  \BibitemOpen
  \bibfield  {author} {\bibinfo {author} {\bibfnamefont {P.}~\bibnamefont
  {Govindaraj}}, \bibinfo {author} {\bibfnamefont {M.}~\bibnamefont
  {Sivasamy}}, \bibinfo {author} {\bibfnamefont {K.}~\bibnamefont {Murugan}},
  \bibinfo {author} {\bibfnamefont {K.}~\bibnamefont {Venugopal}},\ and\
  \bibinfo {author} {\bibfnamefont {P.}~\bibnamefont {Veluswamy}},\ }\bibfield
  {title} {\bibinfo {title} {{Pressure-driven thermoelectric properties of
  defect chalcopyrite structured ZnGa2Te4: Ab initio study}},\ }\href
  {https://doi.org/10.1039/D2RA00805J} {\bibfield  {journal} {\bibinfo
  {journal} {RSC Advances}\ }\textbf {\bibinfo {volume} {12}},\ \bibinfo
  {pages} {12573} (\bibinfo {year} {2022})}\BibitemShut {NoStop}%
\bibitem [{\citenamefont {Qi}\ \emph {et~al.}(2022)\citenamefont {Qi},
  \citenamefont {Qu}, \citenamefont {Liu}, \citenamefont {Qiu}, \citenamefont
  {Li}, \citenamefont {Yue},\ and\ \citenamefont {Guo}}]{qi2022large}%
  \BibitemOpen
  \bibfield  {author} {\bibinfo {author} {\bibfnamefont {H.}~\bibnamefont
  {Qi}}, \bibinfo {author} {\bibfnamefont {T.}~\bibnamefont {Qu}}, \bibinfo
  {author} {\bibfnamefont {Z.}~\bibnamefont {Liu}}, \bibinfo {author}
  {\bibfnamefont {Z.}~\bibnamefont {Qiu}}, \bibinfo {author} {\bibfnamefont
  {C.}~\bibnamefont {Li}}, \bibinfo {author} {\bibfnamefont {S.}~\bibnamefont
  {Yue}},\ and\ \bibinfo {author} {\bibfnamefont {J.}~\bibnamefont {Guo}},\
  }\bibfield  {title} {\bibinfo {title} {{Large enhancement of thermoelectric
  properties of CoSb3 tuned by uniaxial strain}},\ }\href
  {https://doi.org/10.1016/j.jallcom.2022.164404} {\bibfield  {journal}
  {\bibinfo  {journal} {Journal of Alloys and Compounds}\ }\textbf {\bibinfo
  {volume} {908}},\ \bibinfo {pages} {164404} (\bibinfo {year}
  {2022})}\BibitemShut {NoStop}%
\bibitem [{\citenamefont {Xia}\ \emph {et~al.}(2024)\citenamefont {Xia},
  \citenamefont {Zhao}, \citenamefont {Chang}, \citenamefont {Liu},
  \citenamefont {Zhang}, \citenamefont {Zhou}, \citenamefont {Zhao},\ and\
  \citenamefont {Gao}}]{xia2024strain}%
  \BibitemOpen
  \bibfield  {author} {\bibinfo {author} {\bibfnamefont {M.}~\bibnamefont
  {Xia}}, \bibinfo {author} {\bibfnamefont {L.}~\bibnamefont {Zhao}}, \bibinfo
  {author} {\bibfnamefont {Y.}~\bibnamefont {Chang}}, \bibinfo {author}
  {\bibfnamefont {H.}~\bibnamefont {Liu}}, \bibinfo {author} {\bibfnamefont
  {G.}~\bibnamefont {Zhang}}, \bibinfo {author} {\bibfnamefont
  {W.}~\bibnamefont {Zhou}}, \bibinfo {author} {\bibfnamefont {J.}~\bibnamefont
  {Zhao}},\ and\ \bibinfo {author} {\bibfnamefont {J.}~\bibnamefont {Gao}},\
  }\bibfield  {title} {\bibinfo {title} {Strain controlled thermal regulator
  realized in two-dimensional black and blue phosphorene in-plane
  heterostructure},\ }\href {https://doi.org/10.1103/PhysRevB.109.104106}
  {\bibfield  {journal} {\bibinfo  {journal} {Physical Review B}\ }\textbf
  {\bibinfo {volume} {109}},\ \bibinfo {pages} {104106} (\bibinfo {year}
  {2024})}\BibitemShut {NoStop}%
\bibitem [{\citenamefont {Gorai}\ \emph {et~al.}(2017)\citenamefont {Gorai},
  \citenamefont {Stevanovi{\'c}},\ and\ \citenamefont
  {Toberer}}]{gorai2017computationally}%
  \BibitemOpen
  \bibfield  {author} {\bibinfo {author} {\bibfnamefont {P.}~\bibnamefont
  {Gorai}}, \bibinfo {author} {\bibfnamefont {V.}~\bibnamefont
  {Stevanovi{\'c}}},\ and\ \bibinfo {author} {\bibfnamefont {E.~S.}\
  \bibnamefont {Toberer}},\ }\bibfield  {title} {\bibinfo {title}
  {Computationally guided discovery of thermoelectric materials},\ }\href
  {https://doi.org/10.1038/natrevmats.2017.53} {\bibfield  {journal} {\bibinfo
  {journal} {Nature Reviews Materials}\ }\textbf {\bibinfo {volume} {2}},\
  \bibinfo {pages} {1} (\bibinfo {year} {2017})}\BibitemShut {NoStop}%
\bibitem [{\citenamefont {Puligheddu}\ \emph {et~al.}(2019)\citenamefont
  {Puligheddu}, \citenamefont {Xia}, \citenamefont {Chan},\ and\ \citenamefont
  {Galli}}]{puligheddu2019computational}%
  \BibitemOpen
  \bibfield  {author} {\bibinfo {author} {\bibfnamefont {M.}~\bibnamefont
  {Puligheddu}}, \bibinfo {author} {\bibfnamefont {Y.}~\bibnamefont {Xia}},
  \bibinfo {author} {\bibfnamefont {M.}~\bibnamefont {Chan}},\ and\ \bibinfo
  {author} {\bibfnamefont {G.}~\bibnamefont {Galli}},\ }\bibfield  {title}
  {\bibinfo {title} {Computational prediction of lattice thermal conductivity:
  A comparison of molecular dynamics and boltzmann transport approaches},\
  }\href {https://doi.org/10.1103/PhysRevMaterials.3.085401} {\bibfield
  {journal} {\bibinfo  {journal} {Physical Review Materials}\ }\textbf
  {\bibinfo {volume} {3}},\ \bibinfo {pages} {085401} (\bibinfo {year}
  {2019})}\BibitemShut {NoStop}%
\bibitem [{\citenamefont {Xia}\ \emph {et~al.}(2020)\citenamefont {Xia},
  \citenamefont {Hegde}, \citenamefont {Pal}, \citenamefont {Hua},
  \citenamefont {Gaines}, \citenamefont {Patel}, \citenamefont {He},
  \citenamefont {Aykol},\ and\ \citenamefont {Wolverton}}]{xia2020high}%
  \BibitemOpen
  \bibfield  {author} {\bibinfo {author} {\bibfnamefont {Y.}~\bibnamefont
  {Xia}}, \bibinfo {author} {\bibfnamefont {V.~I.}\ \bibnamefont {Hegde}},
  \bibinfo {author} {\bibfnamefont {K.}~\bibnamefont {Pal}}, \bibinfo {author}
  {\bibfnamefont {X.}~\bibnamefont {Hua}}, \bibinfo {author} {\bibfnamefont
  {D.}~\bibnamefont {Gaines}}, \bibinfo {author} {\bibfnamefont
  {S.}~\bibnamefont {Patel}}, \bibinfo {author} {\bibfnamefont
  {J.}~\bibnamefont {He}}, \bibinfo {author} {\bibfnamefont {M.}~\bibnamefont
  {Aykol}},\ and\ \bibinfo {author} {\bibfnamefont {C.}~\bibnamefont
  {Wolverton}},\ }\bibfield  {title} {\bibinfo {title} {High-throughput study
  of lattice thermal conductivity in binary rocksalt and zinc blende compounds
  including higher-order anharmonicity},\ }\href
  {https://doi.org/10.1103/PhysRevX.10.041029} {\bibfield  {journal} {\bibinfo
  {journal} {Physical Review X}\ }\textbf {\bibinfo {volume} {10}},\ \bibinfo
  {pages} {041029} (\bibinfo {year} {2020})}\BibitemShut {NoStop}%
\bibitem [{\citenamefont {He}\ \emph {et~al.}(2022)\citenamefont {He},
  \citenamefont {Xia}, \citenamefont {Lin}, \citenamefont {Pal}, \citenamefont
  {Zhu}, \citenamefont {Kanatzidis},\ and\ \citenamefont
  {Wolverton}}]{he2022accelerated}%
  \BibitemOpen
  \bibfield  {author} {\bibinfo {author} {\bibfnamefont {J.}~\bibnamefont
  {He}}, \bibinfo {author} {\bibfnamefont {Y.}~\bibnamefont {Xia}}, \bibinfo
  {author} {\bibfnamefont {W.}~\bibnamefont {Lin}}, \bibinfo {author}
  {\bibfnamefont {K.}~\bibnamefont {Pal}}, \bibinfo {author} {\bibfnamefont
  {Y.}~\bibnamefont {Zhu}}, \bibinfo {author} {\bibfnamefont {M.~G.}\
  \bibnamefont {Kanatzidis}},\ and\ \bibinfo {author} {\bibfnamefont
  {C.}~\bibnamefont {Wolverton}},\ }\bibfield  {title} {\bibinfo {title}
  {Accelerated discovery and design of ultralow lattice thermal conductivity
  materials using chemical bonding principles},\ }\href
  {https://doi.org/10.1002/adfm.202108532} {\bibfield  {journal} {\bibinfo
  {journal} {Advanced Functional Materials}\ }\textbf {\bibinfo {volume}
  {32}},\ \bibinfo {pages} {2108532} (\bibinfo {year} {2022})}\BibitemShut
  {NoStop}%
\bibitem [{\citenamefont {Xia}\ \emph {et~al.}(2023)\citenamefont {Xia},
  \citenamefont {Gaines}, \citenamefont {He}, \citenamefont {Pal},
  \citenamefont {Li}, \citenamefont {Kanatzidis}, \citenamefont
  {Ozoli{\c{n}}{\v{s}}},\ and\ \citenamefont {Wolverton}}]{xia2023unified}%
  \BibitemOpen
  \bibfield  {author} {\bibinfo {author} {\bibfnamefont {Y.}~\bibnamefont
  {Xia}}, \bibinfo {author} {\bibfnamefont {D.}~\bibnamefont {Gaines}},
  \bibinfo {author} {\bibfnamefont {J.}~\bibnamefont {He}}, \bibinfo {author}
  {\bibfnamefont {K.}~\bibnamefont {Pal}}, \bibinfo {author} {\bibfnamefont
  {Z.}~\bibnamefont {Li}}, \bibinfo {author} {\bibfnamefont {M.~G.}\
  \bibnamefont {Kanatzidis}}, \bibinfo {author} {\bibfnamefont
  {V.}~\bibnamefont {Ozoli{\c{n}}{\v{s}}}},\ and\ \bibinfo {author}
  {\bibfnamefont {C.}~\bibnamefont {Wolverton}},\ }\bibfield  {title} {\bibinfo
  {title} {A unified understanding of minimum lattice thermal conductivity},\
  }\href {https://doi.org/10.1073/pnas.2302541120} {\bibfield  {journal}
  {\bibinfo  {journal} {Proceedings of the National Academy of Sciences}\
  }\textbf {\bibinfo {volume} {120}},\ \bibinfo {pages} {e2302541120} (\bibinfo
  {year} {2023})}\BibitemShut {NoStop}%
\bibitem [{\citenamefont {Ma}\ \emph {et~al.}(2024)\citenamefont {Ma},
  \citenamefont {Xiao},\ and\ \citenamefont {Xie}}]{ma2024multilayer}%
  \BibitemOpen
  \bibfield  {author} {\bibinfo {author} {\bibfnamefont {N.}~\bibnamefont
  {Ma}}, \bibinfo {author} {\bibfnamefont {C.}~\bibnamefont {Xiao}},\ and\
  \bibinfo {author} {\bibfnamefont {Y.}~\bibnamefont {Xie}},\ }\bibfield
  {title} {\bibinfo {title} {Multilayer approach for ultralow lattice thermal
  conductivity in low-dimensional solids},\ }\href
  {https://doi.org/10.1021/accountsmr.3c00089} {\bibfield  {journal} {\bibinfo
  {journal} {Accounts of Materials Research}\ }\textbf {\bibinfo {volume}
  {5}},\ \bibinfo {pages} {286} (\bibinfo {year} {2024})}\BibitemShut {NoStop}%
\bibitem [{\citenamefont {Gaultois}\ \emph {et~al.}(2013)\citenamefont
  {Gaultois}, \citenamefont {Sparks}, \citenamefont {Borg}, \citenamefont
  {Seshadri}, \citenamefont {Bonificio},\ and\ \citenamefont
  {Clarke}}]{gaultois2013data}%
  \BibitemOpen
  \bibfield  {author} {\bibinfo {author} {\bibfnamefont {M.~W.}\ \bibnamefont
  {Gaultois}}, \bibinfo {author} {\bibfnamefont {T.~D.}\ \bibnamefont
  {Sparks}}, \bibinfo {author} {\bibfnamefont {C.~K.}\ \bibnamefont {Borg}},
  \bibinfo {author} {\bibfnamefont {R.}~\bibnamefont {Seshadri}}, \bibinfo
  {author} {\bibfnamefont {W.~D.}\ \bibnamefont {Bonificio}},\ and\ \bibinfo
  {author} {\bibfnamefont {D.~R.}\ \bibnamefont {Clarke}},\ }\bibfield  {title}
  {\bibinfo {title} {Data-driven review of thermoelectric materials:
  performance and resource considerations},\ }\href
  {https://doi.org/10.1021/cm400893e} {\bibfield  {journal} {\bibinfo
  {journal} {Chemistry of Materials}\ }\textbf {\bibinfo {volume} {25}},\
  \bibinfo {pages} {2911} (\bibinfo {year} {2013})}\BibitemShut {NoStop}%
\bibitem [{\citenamefont {Furmanchuk}\ \emph {et~al.}(2018)\citenamefont
  {Furmanchuk}, \citenamefont {Saal}, \citenamefont {Doak}, \citenamefont
  {Olson}, \citenamefont {Choudhary},\ and\ \citenamefont
  {Agrawal}}]{furmanchuk2018prediction}%
  \BibitemOpen
  \bibfield  {author} {\bibinfo {author} {\bibfnamefont {A.}~\bibnamefont
  {Furmanchuk}}, \bibinfo {author} {\bibfnamefont {J.~E.}\ \bibnamefont
  {Saal}}, \bibinfo {author} {\bibfnamefont {J.~W.}\ \bibnamefont {Doak}},
  \bibinfo {author} {\bibfnamefont {G.~B.}\ \bibnamefont {Olson}}, \bibinfo
  {author} {\bibfnamefont {A.}~\bibnamefont {Choudhary}},\ and\ \bibinfo
  {author} {\bibfnamefont {A.}~\bibnamefont {Agrawal}},\ }\bibfield  {title}
  {\bibinfo {title} {Prediction of seebeck coefficient for compounds without
  restriction to fixed stoichiometry: A machine learning approach},\ }\href
  {https://doi.org/10.1002/jcc.25067} {\bibfield  {journal} {\bibinfo
  {journal} {Journal of Computational Chemistry}\ }\textbf {\bibinfo {volume}
  {39}},\ \bibinfo {pages} {191} (\bibinfo {year} {2018})}\BibitemShut
  {NoStop}%
\bibitem [{\citenamefont {Choudhary}\ \emph {et~al.}(2020)\citenamefont
  {Choudhary}, \citenamefont {Garrity},\ and\ \citenamefont
  {Tavazza}}]{choudhary2020data}%
  \BibitemOpen
  \bibfield  {author} {\bibinfo {author} {\bibfnamefont {K.}~\bibnamefont
  {Choudhary}}, \bibinfo {author} {\bibfnamefont {K.~F.}\ \bibnamefont
  {Garrity}},\ and\ \bibinfo {author} {\bibfnamefont {F.}~\bibnamefont
  {Tavazza}},\ }\bibfield  {title} {\bibinfo {title} {{Data-driven discovery of
  3D and 2D thermoelectric materials}},\ }\href
  {https://doi.org/10.1088/1361-648X/aba06b} {\bibfield  {journal} {\bibinfo
  {journal} {Journal of Physics: Condensed Matter}\ }\textbf {\bibinfo {volume}
  {32}},\ \bibinfo {pages} {475501} (\bibinfo {year} {2020})}\BibitemShut
  {NoStop}%
\bibitem [{\citenamefont {Wang}\ \emph {et~al.}(2020)\citenamefont {Wang},
  \citenamefont {Zhang}, \citenamefont {Snoussi},\ and\ \citenamefont
  {Zhang}}]{wang2020machine}%
  \BibitemOpen
  \bibfield  {author} {\bibinfo {author} {\bibfnamefont {T.}~\bibnamefont
  {Wang}}, \bibinfo {author} {\bibfnamefont {C.}~\bibnamefont {Zhang}},
  \bibinfo {author} {\bibfnamefont {H.}~\bibnamefont {Snoussi}},\ and\ \bibinfo
  {author} {\bibfnamefont {G.}~\bibnamefont {Zhang}},\ }\bibfield  {title}
  {\bibinfo {title} {Machine learning approaches for thermoelectric materials
  research},\ }\href {https://doi.org/10.1002/adfm.201906041} {\bibfield
  {journal} {\bibinfo  {journal} {Advanced Functional Materials}\ }\textbf
  {\bibinfo {volume} {30}},\ \bibinfo {pages} {1906041} (\bibinfo {year}
  {2020})}\BibitemShut {NoStop}%
\bibitem [{\citenamefont {Liu}\ \emph {et~al.}(2020)\citenamefont {Liu},
  \citenamefont {Han}, \citenamefont {Cao}, \citenamefont {Zhou}, \citenamefont
  {Sheng},\ and\ \citenamefont {Liu}}]{liu2020high}%
  \BibitemOpen
  \bibfield  {author} {\bibinfo {author} {\bibfnamefont {J.}~\bibnamefont
  {Liu}}, \bibinfo {author} {\bibfnamefont {S.}~\bibnamefont {Han}}, \bibinfo
  {author} {\bibfnamefont {G.}~\bibnamefont {Cao}}, \bibinfo {author}
  {\bibfnamefont {Z.}~\bibnamefont {Zhou}}, \bibinfo {author} {\bibfnamefont
  {C.}~\bibnamefont {Sheng}},\ and\ \bibinfo {author} {\bibfnamefont
  {H.}~\bibnamefont {Liu}},\ }\bibfield  {title} {\bibinfo {title} {A
  high-throughput descriptor for prediction of lattice thermal conductivity of
  half-heusler compounds},\ }\href {https://doi.org/10.1088/1361-6463/ab898e}
  {\bibfield  {journal} {\bibinfo  {journal} {Journal of Physics D: Applied
  Physics}\ }\textbf {\bibinfo {volume} {53}},\ \bibinfo {pages} {315301}
  (\bibinfo {year} {2020})}\BibitemShut {NoStop}%
\bibitem [{\citenamefont {Chen}\ \emph {et~al.}(2021)\citenamefont {Chen},
  \citenamefont {Schmidt}, \citenamefont {Yan}, \citenamefont {Vohra},\ and\
  \citenamefont {Chen}}]{chen2021machine}%
  \BibitemOpen
  \bibfield  {author} {\bibinfo {author} {\bibfnamefont {W.-C.}\ \bibnamefont
  {Chen}}, \bibinfo {author} {\bibfnamefont {J.~N.}\ \bibnamefont {Schmidt}},
  \bibinfo {author} {\bibfnamefont {D.}~\bibnamefont {Yan}}, \bibinfo {author}
  {\bibfnamefont {Y.~K.}\ \bibnamefont {Vohra}},\ and\ \bibinfo {author}
  {\bibfnamefont {C.-C.}\ \bibnamefont {Chen}},\ }\bibfield  {title} {\bibinfo
  {title} {{Machine learning and evolutionary prediction of superhard BCN
  compounds}},\ }\href {https://doi.org/10.1038/s41524-021-00585-7} {\bibfield
  {journal} {\bibinfo  {journal} {npj Computational Materials}\ }\textbf
  {\bibinfo {volume} {7}},\ \bibinfo {pages} {114} (\bibinfo {year}
  {2021})}\BibitemShut {NoStop}%
\bibitem [{\citenamefont {Mbaye}\ \emph {et~al.}(2021)\citenamefont {Mbaye},
  \citenamefont {Pradhan},\ and\ \citenamefont {Bahoura}}]{mbaye2021data}%
  \BibitemOpen
  \bibfield  {author} {\bibinfo {author} {\bibfnamefont {M.~T.}\ \bibnamefont
  {Mbaye}}, \bibinfo {author} {\bibfnamefont {S.~K.}\ \bibnamefont {Pradhan}},\
  and\ \bibinfo {author} {\bibfnamefont {M.}~\bibnamefont {Bahoura}},\
  }\bibfield  {title} {\bibinfo {title} {Data-driven thermoelectric modeling:
  Current challenges and prospects},\ }\bibfield  {journal} {\bibinfo
  {journal} {Journal of Applied Physics}\ }\textbf {\bibinfo {volume} {130}},\
  \href {https://doi.org/10.1063/5.0054532} {10.1063/5.0054532} (\bibinfo
  {year} {2021})\BibitemShut {NoStop}%
\bibitem [{\citenamefont {Han}\ \emph {et~al.}(2021)\citenamefont {Han},
  \citenamefont {Sun}, \citenamefont {Feng}, \citenamefont {Lin},\ and\
  \citenamefont {Lu}}]{han2021machine}%
  \BibitemOpen
  \bibfield  {author} {\bibinfo {author} {\bibfnamefont {G.}~\bibnamefont
  {Han}}, \bibinfo {author} {\bibfnamefont {Y.}~\bibnamefont {Sun}}, \bibinfo
  {author} {\bibfnamefont {Y.}~\bibnamefont {Feng}}, \bibinfo {author}
  {\bibfnamefont {G.}~\bibnamefont {Lin}},\ and\ \bibinfo {author}
  {\bibfnamefont {N.}~\bibnamefont {Lu}},\ }\bibfield  {title} {\bibinfo
  {title} {Machine learning regression guided thermoelectric materials
  discovery--a review},\ }\href {https://doi.org/10.30919/esmm5f451} {\bibfield
   {journal} {\bibinfo  {journal} {ES Materials \& Manufacturing}\ }\textbf
  {\bibinfo {volume} {14}},\ \bibinfo {pages} {20} (\bibinfo {year}
  {2021})}\BibitemShut {NoStop}%
\bibitem [{\citenamefont {An}(2022)}]{american2022machine}%
  \BibitemOpen
  \bibinfo {editor} {\bibfnamefont {Y.}~\bibnamefont {An}},\ ed.,\ \href
  {https://doi.org/10.1021/bk-2022-1416} {\emph {\bibinfo {title} {Machine
  Learning in Materials Informatics: Methods and Applications}}}\ (\bibinfo
  {publisher} {ACS Publications},\ \bibinfo {year} {2022})\BibitemShut
  {NoStop}%
\bibitem [{\citenamefont {Wang}\ \emph
  {et~al.}(2023{\natexlab{a}})\citenamefont {Wang}, \citenamefont {Sheng},
  \citenamefont {Ning}, \citenamefont {Xi}, \citenamefont {Xi}, \citenamefont
  {Qiu}, \citenamefont {Yang},\ and\ \citenamefont {Ke}}]{wang2023critical}%
  \BibitemOpen
  \bibfield  {author} {\bibinfo {author} {\bibfnamefont {X.}~\bibnamefont
  {Wang}}, \bibinfo {author} {\bibfnamefont {Y.}~\bibnamefont {Sheng}},
  \bibinfo {author} {\bibfnamefont {J.}~\bibnamefont {Ning}}, \bibinfo {author}
  {\bibfnamefont {J.}~\bibnamefont {Xi}}, \bibinfo {author} {\bibfnamefont
  {L.}~\bibnamefont {Xi}}, \bibinfo {author} {\bibfnamefont {D.}~\bibnamefont
  {Qiu}}, \bibinfo {author} {\bibfnamefont {J.}~\bibnamefont {Yang}},\ and\
  \bibinfo {author} {\bibfnamefont {X.}~\bibnamefont {Ke}},\ }\bibfield
  {title} {\bibinfo {title} {A critical review of machine learning techniques
  on thermoelectric materials},\ }\href
  {https://doi.org/10.1021/acs.jpclett.2c03073} {\bibfield  {journal} {\bibinfo
   {journal} {The Journal of Physical Chemistry Letters}\ }\textbf {\bibinfo
  {volume} {14}},\ \bibinfo {pages} {1808} (\bibinfo {year}
  {2023}{\natexlab{a}})}\BibitemShut {NoStop}%
\bibitem [{\citenamefont {Wu}\ \emph {et~al.}(2023)\citenamefont {Wu},
  \citenamefont {Li}, \citenamefont {Zeng}, \citenamefont {Zhao}, \citenamefont
  {Xie}, \citenamefont {Zhou}, \citenamefont {Liu},\ and\ \citenamefont
  {Zhang}}]{wu2023machine}%
  \BibitemOpen
  \bibfield  {author} {\bibinfo {author} {\bibfnamefont {C.-W.}\ \bibnamefont
  {Wu}}, \bibinfo {author} {\bibfnamefont {F.}~\bibnamefont {Li}}, \bibinfo
  {author} {\bibfnamefont {Y.-J.}\ \bibnamefont {Zeng}}, \bibinfo {author}
  {\bibfnamefont {H.}~\bibnamefont {Zhao}}, \bibinfo {author} {\bibfnamefont
  {G.}~\bibnamefont {Xie}}, \bibinfo {author} {\bibfnamefont {W.-X.}\
  \bibnamefont {Zhou}}, \bibinfo {author} {\bibfnamefont {Q.}~\bibnamefont
  {Liu}},\ and\ \bibinfo {author} {\bibfnamefont {G.}~\bibnamefont {Zhang}},\
  }\bibfield  {title} {\bibinfo {title} {{Machine learning accelerated design
  of 2D covalent organic frame materials for thermoelectrics}},\ }\href
  {https://doi.org/10.1016/j.apsusc.2023.157947} {\bibfield  {journal}
  {\bibinfo  {journal} {Applied Surface Science}\ }\textbf {\bibinfo {volume}
  {638}},\ \bibinfo {pages} {157947} (\bibinfo {year} {2023})}\BibitemShut
  {NoStop}%
\bibitem [{\citenamefont {Purcell}\ \emph {et~al.}(2023)\citenamefont
  {Purcell}, \citenamefont {Scheffler}, \citenamefont {Ghiringhelli},\ and\
  \citenamefont {Carbogno}}]{purcell2023accelerating}%
  \BibitemOpen
  \bibfield  {author} {\bibinfo {author} {\bibfnamefont {T.~A.}\ \bibnamefont
  {Purcell}}, \bibinfo {author} {\bibfnamefont {M.}~\bibnamefont {Scheffler}},
  \bibinfo {author} {\bibfnamefont {L.~M.}\ \bibnamefont {Ghiringhelli}},\ and\
  \bibinfo {author} {\bibfnamefont {C.}~\bibnamefont {Carbogno}},\ }\bibfield
  {title} {\bibinfo {title} {Accelerating materials-space exploration for
  thermal insulators by mapping materials properties via artificial
  intelligence},\ }\href {https://doi.org/10.1038/s41524-023-01063-y}
  {\bibfield  {journal} {\bibinfo  {journal} {npj Computational Materials}\
  }\textbf {\bibinfo {volume} {9}},\ \bibinfo {pages} {112} (\bibinfo {year}
  {2023})}\BibitemShut {NoStop}%
\bibitem [{\citenamefont {Liu}\ \emph {et~al.}(2024)\citenamefont {Liu},
  \citenamefont {Yao}, \citenamefont {Wang}, \citenamefont {Jin}, \citenamefont
  {Ji}, \citenamefont {Luo}, \citenamefont {Cao}, \citenamefont {Xiong},
  \citenamefont {Sheng}, \citenamefont {Li} \emph {et~al.}}]{liu2024mathub}%
  \BibitemOpen
  \bibfield  {author} {\bibinfo {author} {\bibfnamefont {L.}~\bibnamefont
  {Liu}}, \bibinfo {author} {\bibfnamefont {M.}~\bibnamefont {Yao}}, \bibinfo
  {author} {\bibfnamefont {Y.}~\bibnamefont {Wang}}, \bibinfo {author}
  {\bibfnamefont {Y.}~\bibnamefont {Jin}}, \bibinfo {author} {\bibfnamefont
  {J.}~\bibnamefont {Ji}}, \bibinfo {author} {\bibfnamefont {H.}~\bibnamefont
  {Luo}}, \bibinfo {author} {\bibfnamefont {Y.}~\bibnamefont {Cao}}, \bibinfo
  {author} {\bibfnamefont {Y.}~\bibnamefont {Xiong}}, \bibinfo {author}
  {\bibfnamefont {Y.}~\bibnamefont {Sheng}}, \bibinfo {author} {\bibfnamefont
  {X.}~\bibnamefont {Li}}, \emph {et~al.},\ }\bibfield  {title} {\bibinfo
  {title} {The {MatHub-3d} first-principles repository and the applications on
  thermoelectrics},\ }\href {https://doi.org/10.1002/mgea.21} {\bibfield
  {journal} {\bibinfo  {journal} {Materials Genome Engineering Advances}\
  }\textbf {\bibinfo {volume} {2}},\ \bibinfo {pages} {e21} (\bibinfo {year}
  {2024})}\BibitemShut {NoStop}%
\bibitem [{\citenamefont {Soleimani}\ \emph {et~al.}(2020)\citenamefont
  {Soleimani}, \citenamefont {Zoras}, \citenamefont {Ceranic}, \citenamefont
  {Shahzad},\ and\ \citenamefont {Cui}}]{soleimani2020review}%
  \BibitemOpen
  \bibfield  {author} {\bibinfo {author} {\bibfnamefont {Z.}~\bibnamefont
  {Soleimani}}, \bibinfo {author} {\bibfnamefont {S.}~\bibnamefont {Zoras}},
  \bibinfo {author} {\bibfnamefont {B.}~\bibnamefont {Ceranic}}, \bibinfo
  {author} {\bibfnamefont {S.}~\bibnamefont {Shahzad}},\ and\ \bibinfo {author}
  {\bibfnamefont {Y.}~\bibnamefont {Cui}},\ }\bibfield  {title} {\bibinfo
  {title} {A review on recent developments of thermoelectric materials for
  room-temperature applications},\ }\href
  {https://doi.org/10.1016/j.seta.2019.100604} {\bibfield  {journal} {\bibinfo
  {journal} {Sustainable Energy Technologies and Assessments}\ }\textbf
  {\bibinfo {volume} {37}},\ \bibinfo {pages} {100604} (\bibinfo {year}
  {2020})}\BibitemShut {NoStop}%
\bibitem [{\citenamefont {Gorai}\ \emph {et~al.}(2016)\citenamefont {Gorai},
  \citenamefont {Gao}, \citenamefont {Ortiz}, \citenamefont {Miller},
  \citenamefont {Barnett}, \citenamefont {Mason}, \citenamefont {Lv},
  \citenamefont {Stevanovi{\'c}},\ and\ \citenamefont {Toberer}}]{gorai2016te}%
  \BibitemOpen
  \bibfield  {author} {\bibinfo {author} {\bibfnamefont {P.}~\bibnamefont
  {Gorai}}, \bibinfo {author} {\bibfnamefont {D.}~\bibnamefont {Gao}}, \bibinfo
  {author} {\bibfnamefont {B.}~\bibnamefont {Ortiz}}, \bibinfo {author}
  {\bibfnamefont {S.}~\bibnamefont {Miller}}, \bibinfo {author} {\bibfnamefont
  {S.~A.}\ \bibnamefont {Barnett}}, \bibinfo {author} {\bibfnamefont
  {T.}~\bibnamefont {Mason}}, \bibinfo {author} {\bibfnamefont
  {Q.}~\bibnamefont {Lv}}, \bibinfo {author} {\bibfnamefont {V.}~\bibnamefont
  {Stevanovi{\'c}}},\ and\ \bibinfo {author} {\bibfnamefont {E.~S.}\
  \bibnamefont {Toberer}},\ }\bibfield  {title} {\bibinfo {title} {{TE Design
  Lab: A virtual laboratory for thermoelectric material design}},\ }\href
  {https://doi.org/10.1016/j.commatsci.2015.11.006} {\bibfield  {journal}
  {\bibinfo  {journal} {Computational Materials Science}\ }\textbf {\bibinfo
  {volume} {112}},\ \bibinfo {pages} {368} (\bibinfo {year}
  {2016})}\BibitemShut {NoStop}%
\bibitem [{\citenamefont {Ward}\ \emph {et~al.}(2018)\citenamefont {Ward},
  \citenamefont {Dunn}, \citenamefont {Faghaninia}, \citenamefont {Zimmermann},
  \citenamefont {Bajaj}, \citenamefont {Wang}, \citenamefont {Montoya},
  \citenamefont {Chen}, \citenamefont {Bystrom}, \citenamefont {Dylla} \emph
  {et~al.}}]{ward2018matminer}%
  \BibitemOpen
  \bibfield  {author} {\bibinfo {author} {\bibfnamefont {L.}~\bibnamefont
  {Ward}}, \bibinfo {author} {\bibfnamefont {A.}~\bibnamefont {Dunn}}, \bibinfo
  {author} {\bibfnamefont {A.}~\bibnamefont {Faghaninia}}, \bibinfo {author}
  {\bibfnamefont {N.~E.}\ \bibnamefont {Zimmermann}}, \bibinfo {author}
  {\bibfnamefont {S.}~\bibnamefont {Bajaj}}, \bibinfo {author} {\bibfnamefont
  {Q.}~\bibnamefont {Wang}}, \bibinfo {author} {\bibfnamefont {J.}~\bibnamefont
  {Montoya}}, \bibinfo {author} {\bibfnamefont {J.}~\bibnamefont {Chen}},
  \bibinfo {author} {\bibfnamefont {K.}~\bibnamefont {Bystrom}}, \bibinfo
  {author} {\bibfnamefont {M.}~\bibnamefont {Dylla}}, \emph {et~al.},\
  }\bibfield  {title} {\bibinfo {title} {Matminer: An open source toolkit for
  materials data mining},\ }\href
  {https://doi.org/10.1016/j.commatsci.2018.05.018} {\bibfield  {journal}
  {\bibinfo  {journal} {Computational Materials Science}\ }\textbf {\bibinfo
  {volume} {152}},\ \bibinfo {pages} {60} (\bibinfo {year} {2018})}\BibitemShut
  {NoStop}%
\bibitem [{\citenamefont {Chen}\ \emph {et~al.}(2019)\citenamefont {Chen},
  \citenamefont {Tran}, \citenamefont {Batra}, \citenamefont {Kim},\ and\
  \citenamefont {Ramprasad}}]{chen2019machine}%
  \BibitemOpen
  \bibfield  {author} {\bibinfo {author} {\bibfnamefont {L.}~\bibnamefont
  {Chen}}, \bibinfo {author} {\bibfnamefont {H.}~\bibnamefont {Tran}}, \bibinfo
  {author} {\bibfnamefont {R.}~\bibnamefont {Batra}}, \bibinfo {author}
  {\bibfnamefont {C.}~\bibnamefont {Kim}},\ and\ \bibinfo {author}
  {\bibfnamefont {R.}~\bibnamefont {Ramprasad}},\ }\bibfield  {title} {\bibinfo
  {title} {Machine learning models for the lattice thermal conductivity
  prediction of inorganic materials},\ }\href
  {https://doi.org/10.1016/j.commatsci.2019.109155} {\bibfield  {journal}
  {\bibinfo  {journal} {Computational Materials Science}\ }\textbf {\bibinfo
  {volume} {170}},\ \bibinfo {pages} {109155} (\bibinfo {year}
  {2019})}\BibitemShut {NoStop}%
\bibitem [{\citenamefont {Callaway}(1959)}]{callaway1959model}%
  \BibitemOpen
  \bibfield  {author} {\bibinfo {author} {\bibfnamefont {J.}~\bibnamefont
  {Callaway}},\ }\bibfield  {title} {\bibinfo {title} {Model for lattice
  thermal conductivity at low temperatures},\ }\href
  {https://doi.org/10.1103/PhysRev.113.1046} {\bibfield  {journal} {\bibinfo
  {journal} {Physical Review}\ }\textbf {\bibinfo {volume} {113}},\ \bibinfo
  {pages} {1046} (\bibinfo {year} {1959})}\BibitemShut {NoStop}%
\bibitem [{\citenamefont {Slack}(1973)}]{slack1973nonmetallic}%
  \BibitemOpen
  \bibfield  {author} {\bibinfo {author} {\bibfnamefont {G.~A.}\ \bibnamefont
  {Slack}},\ }\bibfield  {title} {\bibinfo {title} {Nonmetallic crystals with
  high thermal conductivity},\ }\href
  {https://doi.org/10.1016/0022-3697(73)90092-9} {\bibfield  {journal}
  {\bibinfo  {journal} {Journal of Physics and Chemistry of Solids}\ }\textbf
  {\bibinfo {volume} {34}},\ \bibinfo {pages} {321} (\bibinfo {year}
  {1973})}\BibitemShut {NoStop}%
\bibitem [{\citenamefont {Smith}\ \emph {et~al.}(2023)\citenamefont {Smith},
  \citenamefont {Harris}, \citenamefont {Camata}, \citenamefont {Yan},\ and\
  \citenamefont {Chen}}]{smith2023machine}%
  \BibitemOpen
  \bibfield  {author} {\bibinfo {author} {\bibfnamefont {A.~D.}\ \bibnamefont
  {Smith}}, \bibinfo {author} {\bibfnamefont {S.~B.}\ \bibnamefont {Harris}},
  \bibinfo {author} {\bibfnamefont {R.~P.}\ \bibnamefont {Camata}}, \bibinfo
  {author} {\bibfnamefont {D.}~\bibnamefont {Yan}},\ and\ \bibinfo {author}
  {\bibfnamefont {C.-C.}\ \bibnamefont {Chen}},\ }\bibfield  {title} {\bibinfo
  {title} {Machine learning the relationship between debye temperature and
  superconducting transition temperature},\ }\href
  {https://doi.org/10.1103/PhysRevB.108.174514} {\bibfield  {journal} {\bibinfo
   {journal} {Physical Review B}\ }\textbf {\bibinfo {volume} {108}},\ \bibinfo
  {pages} {174514} (\bibinfo {year} {2023})}\BibitemShut {NoStop}%
\bibitem [{\citenamefont {Ho}(1998)}]{ho1998random}%
  \BibitemOpen
  \bibfield  {author} {\bibinfo {author} {\bibfnamefont {T.~K.}\ \bibnamefont
  {Ho}},\ }\bibfield  {title} {\bibinfo {title} {The random subspace method for
  constructing decision forests},\ }\href {https://doi.org/10.1109/34.709601}
  {\bibfield  {journal} {\bibinfo  {journal} {IEEE Transactions on Pattern
  Analysis and Machine Intelligence}\ }\textbf {\bibinfo {volume} {20}},\
  \bibinfo {pages} {832} (\bibinfo {year} {1998})}\BibitemShut {NoStop}%
\bibitem [{\citenamefont {Breiman}(2001)}]{breiman2001random}%
  \BibitemOpen
  \bibfield  {author} {\bibinfo {author} {\bibfnamefont {L.}~\bibnamefont
  {Breiman}},\ }\bibfield  {title} {\bibinfo {title} {Random forests},\ }\href
  {https://doi.org/10.1023/A:1010933404324} {\bibfield  {journal} {\bibinfo
  {journal} {Machine Learning}\ }\textbf {\bibinfo {volume} {45}},\ \bibinfo
  {pages} {5} (\bibinfo {year} {2001})}\BibitemShut {NoStop}%
\bibitem [{\citenamefont {Pedregosa}\ \emph {et~al.}(2011)\citenamefont
  {Pedregosa}, \citenamefont {Varoquaux}, \citenamefont {Gramfort},
  \citenamefont {Michel}, \citenamefont {Thirion}, \citenamefont {Grisel},
  \citenamefont {Blondel}, \citenamefont {Prettenhofer}, \citenamefont {Weiss},
  \citenamefont {Dubourg} \emph {et~al.}}]{pedregosa2011scikit}%
  \BibitemOpen
  \bibfield  {author} {\bibinfo {author} {\bibfnamefont {F.}~\bibnamefont
  {Pedregosa}}, \bibinfo {author} {\bibfnamefont {G.}~\bibnamefont
  {Varoquaux}}, \bibinfo {author} {\bibfnamefont {A.}~\bibnamefont {Gramfort}},
  \bibinfo {author} {\bibfnamefont {V.}~\bibnamefont {Michel}}, \bibinfo
  {author} {\bibfnamefont {B.}~\bibnamefont {Thirion}}, \bibinfo {author}
  {\bibfnamefont {O.}~\bibnamefont {Grisel}}, \bibinfo {author} {\bibfnamefont
  {M.}~\bibnamefont {Blondel}}, \bibinfo {author} {\bibfnamefont
  {P.}~\bibnamefont {Prettenhofer}}, \bibinfo {author} {\bibfnamefont
  {R.}~\bibnamefont {Weiss}}, \bibinfo {author} {\bibfnamefont
  {V.}~\bibnamefont {Dubourg}}, \emph {et~al.},\ }\bibfield  {title} {\bibinfo
  {title} {Scikit-learn: Machine learning in python},\ }\href
  {https://doi.org/http://jmlr.org/papers/v12/pedregosa11a.html} {\bibfield
  {journal} {\bibinfo  {journal} {Journal of Machine Learning Research}\
  }\textbf {\bibinfo {volume} {12}},\ \bibinfo {pages} {2825} (\bibinfo {year}
  {2011})}\BibitemShut {NoStop}%
\bibitem [{\citenamefont {Kresse}\ and\ \citenamefont
  {Furthm{\"u}ller}(1996{\natexlab{a}})}]{kresse1996efficiency}%
  \BibitemOpen
  \bibfield  {author} {\bibinfo {author} {\bibfnamefont {G.}~\bibnamefont
  {Kresse}}\ and\ \bibinfo {author} {\bibfnamefont {J.}~\bibnamefont
  {Furthm{\"u}ller}},\ }\bibfield  {title} {\bibinfo {title} {Efficiency of
  ab-initio total energy calculations for metals and semiconductors using a
  plane-wave basis set},\ }\href {https://doi.org/10.1016/0927-0256(96)00008-0}
  {\bibfield  {journal} {\bibinfo  {journal} {Computational Materials Science}\
  }\textbf {\bibinfo {volume} {6}},\ \bibinfo {pages} {15} (\bibinfo {year}
  {1996}{\natexlab{a}})}\BibitemShut {NoStop}%
\bibitem [{\citenamefont {Kresse}\ and\ \citenamefont
  {Furthm{\"u}ller}(1996{\natexlab{b}})}]{kresse1996efficient}%
  \BibitemOpen
  \bibfield  {author} {\bibinfo {author} {\bibfnamefont {G.}~\bibnamefont
  {Kresse}}\ and\ \bibinfo {author} {\bibfnamefont {J.}~\bibnamefont
  {Furthm{\"u}ller}},\ }\bibfield  {title} {\bibinfo {title} {Efficient
  iterative schemes for ab initio total-energy calculations using a plane-wave
  basis set},\ }\href {https://doi.org/10.1103/PhysRevB.54.11169} {\bibfield
  {journal} {\bibinfo  {journal} {Physical Review B}\ }\textbf {\bibinfo
  {volume} {54}},\ \bibinfo {pages} {11169} (\bibinfo {year}
  {1996}{\natexlab{b}})}\BibitemShut {NoStop}%
\bibitem [{\citenamefont {Bl{\"o}chl}(1994)}]{blochl1994projector}%
  \BibitemOpen
  \bibfield  {author} {\bibinfo {author} {\bibfnamefont {P.~E.}\ \bibnamefont
  {Bl{\"o}chl}},\ }\bibfield  {title} {\bibinfo {title} {Projector
  augmented-wave method},\ }\href {https://doi.org/10.1103/PhysRevB.50.17953}
  {\bibfield  {journal} {\bibinfo  {journal} {Physical Review B}\ }\textbf
  {\bibinfo {volume} {50}},\ \bibinfo {pages} {17953} (\bibinfo {year}
  {1994})}\BibitemShut {NoStop}%
\bibitem [{\citenamefont {Kresse}\ and\ \citenamefont
  {Joubert}(1999)}]{kresse1999ultrasoft}%
  \BibitemOpen
  \bibfield  {author} {\bibinfo {author} {\bibfnamefont {G.}~\bibnamefont
  {Kresse}}\ and\ \bibinfo {author} {\bibfnamefont {D.}~\bibnamefont
  {Joubert}},\ }\bibfield  {title} {\bibinfo {title} {From ultrasoft
  pseudopotentials to the projector augmented-wave method},\ }\href
  {https://doi.org/10.1103/PhysRevB.59.1758} {\bibfield  {journal} {\bibinfo
  {journal} {Physical Review B}\ }\textbf {\bibinfo {volume} {59}},\ \bibinfo
  {pages} {1758} (\bibinfo {year} {1999})}\BibitemShut {NoStop}%
\bibitem [{\citenamefont {Perdew}\ \emph {et~al.}(1996)\citenamefont {Perdew},
  \citenamefont {Burke},\ and\ \citenamefont
  {Ernzerhof}}]{perdew1996generalized}%
  \BibitemOpen
  \bibfield  {author} {\bibinfo {author} {\bibfnamefont {J.~P.}\ \bibnamefont
  {Perdew}}, \bibinfo {author} {\bibfnamefont {K.}~\bibnamefont {Burke}},\ and\
  \bibinfo {author} {\bibfnamefont {M.}~\bibnamefont {Ernzerhof}},\ }\bibfield
  {title} {\bibinfo {title} {Generalized gradient approximation made simple},\
  }\href {https://doi.org/10.1103/PhysRevLett.77.3865} {\bibfield  {journal}
  {\bibinfo  {journal} {Physical Review Letters}\ }\textbf {\bibinfo {volume}
  {77}},\ \bibinfo {pages} {3865} (\bibinfo {year} {1996})}\BibitemShut
  {NoStop}%
\bibitem [{\citenamefont {Monkhorst}\ and\ \citenamefont
  {Pack}(1976)}]{monkhorst1976special}%
  \BibitemOpen
  \bibfield  {author} {\bibinfo {author} {\bibfnamefont {H.~J.}\ \bibnamefont
  {Monkhorst}}\ and\ \bibinfo {author} {\bibfnamefont {J.~D.}\ \bibnamefont
  {Pack}},\ }\bibfield  {title} {\bibinfo {title} {Special points for
  brillouin-zone integrations},\ }\href
  {https://doi.org/10.1103/PhysRevB.13.5188} {\bibfield  {journal} {\bibinfo
  {journal} {Physical Review B}\ }\textbf {\bibinfo {volume} {13}},\ \bibinfo
  {pages} {5188} (\bibinfo {year} {1976})}\BibitemShut {NoStop}%
\bibitem [{\citenamefont {Madsen}\ \emph {et~al.}(2018)\citenamefont {Madsen},
  \citenamefont {Carrete},\ and\ \citenamefont
  {Verstraete}}]{madsen2018boltztrap2}%
  \BibitemOpen
  \bibfield  {author} {\bibinfo {author} {\bibfnamefont {G.~K.}\ \bibnamefont
  {Madsen}}, \bibinfo {author} {\bibfnamefont {J.}~\bibnamefont {Carrete}},\
  and\ \bibinfo {author} {\bibfnamefont {M.~J.}\ \bibnamefont {Verstraete}},\
  }\bibfield  {title} {\bibinfo {title} {{BoltzTraP2, a program for
  interpolating band structures and calculating semi-classical transport
  coefficients}},\ }\href {https://doi.org/10.1016/j.cpc.2018.05.010}
  {\bibfield  {journal} {\bibinfo  {journal} {Computer Physics Communications}\
  }\textbf {\bibinfo {volume} {231}},\ \bibinfo {pages} {140} (\bibinfo {year}
  {2018})}\BibitemShut {NoStop}%
\bibitem [{\citenamefont {Togo}\ and\ \citenamefont {Tanaka}(2015)}]{phonopy}%
  \BibitemOpen
  \bibfield  {author} {\bibinfo {author} {\bibfnamefont {A.}~\bibnamefont
  {Togo}}\ and\ \bibinfo {author} {\bibfnamefont {I.}~\bibnamefont {Tanaka}},\
  }\bibfield  {title} {\bibinfo {title} {First principles phonon calculations
  in materials science},\ }\href
  {https://doi.org/10.1016/j.scriptamat.2015.07.021} {\bibfield  {journal}
  {\bibinfo  {journal} {Scripta Materialia}\ }\textbf {\bibinfo {volume}
  {108}},\ \bibinfo {pages} {1} (\bibinfo {year} {2015})}\BibitemShut {NoStop}%
\bibitem [{\citenamefont {Togo}\ \emph {et~al.}(2015)\citenamefont {Togo},
  \citenamefont {Chaput},\ and\ \citenamefont {Tanaka}}]{phono3py}%
  \BibitemOpen
  \bibfield  {author} {\bibinfo {author} {\bibfnamefont {A.}~\bibnamefont
  {Togo}}, \bibinfo {author} {\bibfnamefont {L.}~\bibnamefont {Chaput}},\ and\
  \bibinfo {author} {\bibfnamefont {I.}~\bibnamefont {Tanaka}},\ }\bibfield
  {title} {\bibinfo {title} {Distributions of phonon lifetimes in brillouin
  zones},\ }\href {https://doi.org/10.1103/PhysRevB.91.094306} {\bibfield
  {journal} {\bibinfo  {journal} {Physical Review B}\ }\textbf {\bibinfo
  {volume} {91}},\ \bibinfo {pages} {094306} (\bibinfo {year}
  {2015})}\BibitemShut {NoStop}%
\bibitem [{\citenamefont {Togo}\ \emph {et~al.}(2023)\citenamefont {Togo},
  \citenamefont {Chaput}, \citenamefont {Tadano},\ and\ \citenamefont
  {Tanaka}}]{phonopy-phono3py-JPCM}%
  \BibitemOpen
  \bibfield  {author} {\bibinfo {author} {\bibfnamefont {A.}~\bibnamefont
  {Togo}}, \bibinfo {author} {\bibfnamefont {L.}~\bibnamefont {Chaput}},
  \bibinfo {author} {\bibfnamefont {T.}~\bibnamefont {Tadano}},\ and\ \bibinfo
  {author} {\bibfnamefont {I.}~\bibnamefont {Tanaka}},\ }\bibfield  {title}
  {\bibinfo {title} {Implementation strategies in phonopy and phono3py},\
  }\href {https://doi.org/10.1088/1361-648X/acd831} {\bibfield  {journal}
  {\bibinfo  {journal} {Journal of Physics: Condensed Matter}\ }\textbf
  {\bibinfo {volume} {35}},\ \bibinfo {pages} {353001} (\bibinfo {year}
  {2023})}\BibitemShut {NoStop}%
\bibitem [{\citenamefont {Momma}\ and\ \citenamefont
  {Izumi}(2011)}]{momma2011vesta}%
  \BibitemOpen
  \bibfield  {author} {\bibinfo {author} {\bibfnamefont {K.}~\bibnamefont
  {Momma}}\ and\ \bibinfo {author} {\bibfnamefont {F.}~\bibnamefont {Izumi}},\
  }\bibfield  {title} {\bibinfo {title} {{VESTA 3 for three-dimensional
  visualization of crystal, volumetric and morphology data}},\ }\href
  {https://doi.org/10.1107/S0021889811038970} {\bibfield  {journal} {\bibinfo
  {journal} {Journal of Applied Crystallography}\ }\textbf {\bibinfo {volume}
  {44}},\ \bibinfo {pages} {1272} (\bibinfo {year} {2011})}\BibitemShut
  {NoStop}%
\bibitem [{Per()}]{PeriodicTrendPlotter}%
  \BibitemOpen
  \href@noop {} {\bibinfo {title} {Periodic trend plotter}},\ \bibinfo
  {howpublished} {\url{https://github.com/Andrew-S-Rosen/periodic_trends}},\
  \bibinfo {note} {accessed: 2024-08-08}\BibitemShut {NoStop}%
\bibitem [{\citenamefont {Dutta}\ \emph {et~al.}(2021)\citenamefont {Dutta},
  \citenamefont {Samanta}, \citenamefont {Ghosh}, \citenamefont {Voneshen},\
  and\ \citenamefont {Biswas}}]{dutta2021evidence}%
  \BibitemOpen
  \bibfield  {author} {\bibinfo {author} {\bibfnamefont {M.}~\bibnamefont
  {Dutta}}, \bibinfo {author} {\bibfnamefont {M.}~\bibnamefont {Samanta}},
  \bibinfo {author} {\bibfnamefont {T.}~\bibnamefont {Ghosh}}, \bibinfo
  {author} {\bibfnamefont {D.~J.}\ \bibnamefont {Voneshen}},\ and\ \bibinfo
  {author} {\bibfnamefont {K.}~\bibnamefont {Biswas}},\ }\bibfield  {title}
  {\bibinfo {title} {{Evidence of highly anharmonic soft lattice vibrations in
  a Zintl rattler}},\ }\href {https://doi.org/10.1002/ange.202013923}
  {\bibfield  {journal} {\bibinfo  {journal} {Angewandte Chemie}\ }\textbf
  {\bibinfo {volume} {133}},\ \bibinfo {pages} {4305} (\bibinfo {year}
  {2021})}\BibitemShut {NoStop}%
\bibitem [{\citenamefont {Chen}\ \emph {et~al.}(2018)\citenamefont {Chen},
  \citenamefont {Zhang},\ and\ \citenamefont {Pei}}]{chen2018manipulation}%
  \BibitemOpen
  \bibfield  {author} {\bibinfo {author} {\bibfnamefont {Z.}~\bibnamefont
  {Chen}}, \bibinfo {author} {\bibfnamefont {X.}~\bibnamefont {Zhang}},\ and\
  \bibinfo {author} {\bibfnamefont {Y.}~\bibnamefont {Pei}},\ }\bibfield
  {title} {\bibinfo {title} {Manipulation of phonon transport in
  thermoelectrics},\ }\href {https://doi.org/10.1002/adma.201705617} {\bibfield
   {journal} {\bibinfo  {journal} {Advanced Materials}\ }\textbf {\bibinfo
  {volume} {30}},\ \bibinfo {pages} {1705617} (\bibinfo {year}
  {2018})}\BibitemShut {NoStop}%
\bibitem [{\citenamefont {Chang}\ and\ \citenamefont
  {Zhao}(2018)}]{chang2018anharmoncity}%
  \BibitemOpen
  \bibfield  {author} {\bibinfo {author} {\bibfnamefont {C.}~\bibnamefont
  {Chang}}\ and\ \bibinfo {author} {\bibfnamefont {L.-D.}\ \bibnamefont
  {Zhao}},\ }\bibfield  {title} {\bibinfo {title} {Anharmoncity and low thermal
  conductivity in thermoelectrics},\ }\href
  {https://doi.org/10.1016/j.mtphys.2018.02.005} {\bibfield  {journal}
  {\bibinfo  {journal} {Materials Today Physics}\ }\textbf {\bibinfo {volume}
  {4}},\ \bibinfo {pages} {50} (\bibinfo {year} {2018})}\BibitemShut {NoStop}%
\bibitem [{\citenamefont {Toberer}\ \emph {et~al.}(2010)\citenamefont
  {Toberer}, \citenamefont {May},\ and\ \citenamefont
  {Snyder}}]{toberer2010zintl}%
  \BibitemOpen
  \bibfield  {author} {\bibinfo {author} {\bibfnamefont {E.~S.}\ \bibnamefont
  {Toberer}}, \bibinfo {author} {\bibfnamefont {A.~F.}\ \bibnamefont {May}},\
  and\ \bibinfo {author} {\bibfnamefont {G.~J.}\ \bibnamefont {Snyder}},\
  }\bibfield  {title} {\bibinfo {title} {Zintl chemistry for designing high
  efficiency thermoelectric materials},\ }\href
  {https://doi.org/10.1021/cm901956r} {\bibfield  {journal} {\bibinfo
  {journal} {Chemistry of Materials}\ }\textbf {\bibinfo {volume} {22}},\
  \bibinfo {pages} {624} (\bibinfo {year} {2010})}\BibitemShut {NoStop}%
\bibitem [{\citenamefont {Zevalkink}\ \emph {et~al.}(2012)\citenamefont
  {Zevalkink}, \citenamefont {Zeier}, \citenamefont {Pomrehn}, \citenamefont
  {Schechtel}, \citenamefont {Tremel},\ and\ \citenamefont
  {Snyder}}]{zevalkink2012thermoelectric}%
  \BibitemOpen
  \bibfield  {author} {\bibinfo {author} {\bibfnamefont {A.}~\bibnamefont
  {Zevalkink}}, \bibinfo {author} {\bibfnamefont {W.~G.}\ \bibnamefont
  {Zeier}}, \bibinfo {author} {\bibfnamefont {G.}~\bibnamefont {Pomrehn}},
  \bibinfo {author} {\bibfnamefont {E.}~\bibnamefont {Schechtel}}, \bibinfo
  {author} {\bibfnamefont {W.}~\bibnamefont {Tremel}},\ and\ \bibinfo {author}
  {\bibfnamefont {G.~J.}\ \bibnamefont {Snyder}},\ }\bibfield  {title}
  {\bibinfo {title} {{Thermoelectric properties of Sr3GaSb3 - a chain-forming
  Zintl compound}},\ }\href {https://doi.org/10.1039/C2EE22378C} {\bibfield
  {journal} {\bibinfo  {journal} {Energy \& Environmental Science}\ }\textbf
  {\bibinfo {volume} {5}},\ \bibinfo {pages} {9121} (\bibinfo {year}
  {2012})}\BibitemShut {NoStop}%
\bibitem [{\citenamefont {Ding}\ \emph {et~al.}(2018)\citenamefont {Ding},
  \citenamefont {He}, \citenamefont {Cheng}, \citenamefont {Wang},\ and\
  \citenamefont {Li}}]{ding2018low}%
  \BibitemOpen
  \bibfield  {author} {\bibinfo {author} {\bibfnamefont {G.}~\bibnamefont
  {Ding}}, \bibinfo {author} {\bibfnamefont {J.}~\bibnamefont {He}}, \bibinfo
  {author} {\bibfnamefont {Z.}~\bibnamefont {Cheng}}, \bibinfo {author}
  {\bibfnamefont {X.}~\bibnamefont {Wang}},\ and\ \bibinfo {author}
  {\bibfnamefont {S.}~\bibnamefont {Li}},\ }\bibfield  {title} {\bibinfo
  {title} {{Low lattice thermal conductivity and promising thermoelectric
  figure of merit of Zintl type TlInTe2}},\ }\href
  {https://doi.org/10.1039/C8TC03492C} {\bibfield  {journal} {\bibinfo
  {journal} {Journal of Materials Chemistry C}\ }\textbf {\bibinfo {volume}
  {6}},\ \bibinfo {pages} {13269} (\bibinfo {year} {2018})}\BibitemShut
  {NoStop}%
\bibitem [{\citenamefont {Yin}\ \emph {et~al.}(2019)\citenamefont {Yin},
  \citenamefont {Baskaran},\ and\ \citenamefont {Tiwari}}]{yin2019review}%
  \BibitemOpen
  \bibfield  {author} {\bibinfo {author} {\bibfnamefont {Y.}~\bibnamefont
  {Yin}}, \bibinfo {author} {\bibfnamefont {K.}~\bibnamefont {Baskaran}},\ and\
  \bibinfo {author} {\bibfnamefont {A.}~\bibnamefont {Tiwari}},\ }\bibfield
  {title} {\bibinfo {title} {A review of strategies for developing promising
  thermoelectric materials by controlling thermal conduction},\ }\href
  {https://doi.org/10.1002/pssa.201800904} {\bibfield  {journal} {\bibinfo
  {journal} {Physica Status Solidi (A)}\ }\textbf {\bibinfo {volume} {216}},\
  \bibinfo {pages} {1800904} (\bibinfo {year} {2019})}\BibitemShut {NoStop}%
\bibitem [{\citenamefont {Cai}\ \emph {et~al.}(2019)\citenamefont {Cai},
  \citenamefont {Hu}, \citenamefont {Zhuang},\ and\ \citenamefont
  {Li}}]{cai2019promising}%
  \BibitemOpen
  \bibfield  {author} {\bibinfo {author} {\bibfnamefont {B.}~\bibnamefont
  {Cai}}, \bibinfo {author} {\bibfnamefont {H.}~\bibnamefont {Hu}}, \bibinfo
  {author} {\bibfnamefont {H.-L.}\ \bibnamefont {Zhuang}},\ and\ \bibinfo
  {author} {\bibfnamefont {J.-F.}\ \bibnamefont {Li}},\ }\bibfield  {title}
  {\bibinfo {title} {Promising materials for thermoelectric applications},\
  }\href {https://doi.org/10.1016/j.jallcom.2019.07.147} {\bibfield  {journal}
  {\bibinfo  {journal} {Journal of Alloys and Compounds}\ }\textbf {\bibinfo
  {volume} {806}},\ \bibinfo {pages} {471} (\bibinfo {year}
  {2019})}\BibitemShut {NoStop}%
\bibitem [{\citenamefont {Guo}\ \emph {et~al.}(2021)\citenamefont {Guo},
  \citenamefont {Weng}, \citenamefont {Jiang}, \citenamefont {Zhu},
  \citenamefont {Li}, \citenamefont {Yuan}, \citenamefont {Yang}, \citenamefont
  {Zhang}, \citenamefont {Luo}, \citenamefont {Grin} \emph
  {et~al.}}]{guo2021unveiling}%
  \BibitemOpen
  \bibfield  {author} {\bibinfo {author} {\bibfnamefont {K.}~\bibnamefont
  {Guo}}, \bibinfo {author} {\bibfnamefont {T.}~\bibnamefont {Weng}}, \bibinfo
  {author} {\bibfnamefont {Y.}~\bibnamefont {Jiang}}, \bibinfo {author}
  {\bibfnamefont {Y.}~\bibnamefont {Zhu}}, \bibinfo {author} {\bibfnamefont
  {H.}~\bibnamefont {Li}}, \bibinfo {author} {\bibfnamefont {S.}~\bibnamefont
  {Yuan}}, \bibinfo {author} {\bibfnamefont {J.}~\bibnamefont {Yang}}, \bibinfo
  {author} {\bibfnamefont {J.}~\bibnamefont {Zhang}}, \bibinfo {author}
  {\bibfnamefont {J.}~\bibnamefont {Luo}}, \bibinfo {author} {\bibfnamefont
  {Y.}~\bibnamefont {Grin}}, \emph {et~al.},\ }\bibfield  {title} {\bibinfo
  {title} {Unveiling the origins of low lattice thermal conductivity in
  122-phase zintl compounds},\ }\href
  {https://doi.org/10.1016/j.mtphys.2021.100480} {\bibfield  {journal}
  {\bibinfo  {journal} {Materials Today Physics}\ }\textbf {\bibinfo {volume}
  {21}},\ \bibinfo {pages} {100480} (\bibinfo {year} {2021})}\BibitemShut
  {NoStop}%
\bibitem [{\citenamefont {Wang}\ \emph
  {et~al.}(2023{\natexlab{b}})\citenamefont {Wang}, \citenamefont {Zhang},\
  and\ \citenamefont {Wang}}]{wang2023acoustic}%
  \BibitemOpen
  \bibfield  {author} {\bibinfo {author} {\bibfnamefont {S.-F.}\ \bibnamefont
  {Wang}}, \bibinfo {author} {\bibfnamefont {J.-R.}\ \bibnamefont {Zhang}},\
  and\ \bibinfo {author} {\bibfnamefont {F.-W.}\ \bibnamefont {Wang}},\
  }\bibfield  {title} {\bibinfo {title} {{Acoustic phonon softening enhances
  phonon scattering in Zintl-phase II-IV compounds}},\ }\href
  {https://doi.org/10.1103/PhysRevB.108.235213} {\bibfield  {journal} {\bibinfo
   {journal} {Physical Review B}\ }\textbf {\bibinfo {volume} {108}},\ \bibinfo
  {pages} {235213} (\bibinfo {year} {2023}{\natexlab{b}})}\BibitemShut
  {NoStop}%
\bibitem [{\citenamefont {Tran{\aa}s}\ \emph {et~al.}(2023)\citenamefont
  {Tran{\aa}s}, \citenamefont {L{\o}vvik},\ and\ \citenamefont
  {Berland}}]{tranaas2023lattice}%
  \BibitemOpen
  \bibfield  {author} {\bibinfo {author} {\bibfnamefont {R.}~\bibnamefont
  {Tran{\aa}s}}, \bibinfo {author} {\bibfnamefont {O.~M.}\ \bibnamefont
  {L{\o}vvik}},\ and\ \bibinfo {author} {\bibfnamefont {K.}~\bibnamefont
  {Berland}},\ }\bibfield  {title} {\bibinfo {title} {Lattice thermal
  conductivity from first principles and active learning with gaussian process
  regression},\ }\bibfield  {journal} {\bibinfo  {journal} {arXiv:2309.06786}\
  }\href {https://doi.org/10.48550/arXiv.2309.06786}
  {10.48550/arXiv.2309.06786} (\bibinfo {year} {2023})\BibitemShut {NoStop}%
\bibitem [{\citenamefont {Pandey}\ and\ \citenamefont
  {Singh}(2015)}]{pandey2015high}%
  \BibitemOpen
  \bibfield  {author} {\bibinfo {author} {\bibfnamefont {T.}~\bibnamefont
  {Pandey}}\ and\ \bibinfo {author} {\bibfnamefont {A.~K.}\ \bibnamefont
  {Singh}},\ }\bibfield  {title} {\bibinfo {title} {{High thermopower and ultra
  low thermal conductivity in Cd-based Zintl phase compounds}},\ }\href
  {https://doi.org/10.1039/C5CP02344K} {\bibfield  {journal} {\bibinfo
  {journal} {Physical Chemistry Chemical Physics}\ }\textbf {\bibinfo {volume}
  {17}},\ \bibinfo {pages} {16917} (\bibinfo {year} {2015})}\BibitemShut
  {NoStop}%
\bibitem [{\citenamefont {Zhang}\ \emph {et~al.}(2022)\citenamefont {Zhang},
  \citenamefont {He}, \citenamefont {Jiang}, \citenamefont {Xia}, \citenamefont
  {Gao},\ and\ \citenamefont {Huang}}]{zhang2022remarkable}%
  \BibitemOpen
  \bibfield  {author} {\bibinfo {author} {\bibfnamefont {J.}~\bibnamefont
  {Zhang}}, \bibinfo {author} {\bibfnamefont {D.}~\bibnamefont {He}}, \bibinfo
  {author} {\bibfnamefont {H.}~\bibnamefont {Jiang}}, \bibinfo {author}
  {\bibfnamefont {X.}~\bibnamefont {Xia}}, \bibinfo {author} {\bibfnamefont
  {Y.}~\bibnamefont {Gao}},\ and\ \bibinfo {author} {\bibfnamefont
  {Z.}~\bibnamefont {Huang}},\ }\bibfield  {title} {\bibinfo {title}
  {{Remarkable Thermoelectric Performance in K2CdPb Crystals with 1D Building
  Blocks via Structure Particularity and Bond Heterogeneity}},\ }\href
  {https://doi.org/10.1021/acsaem.2c00484} {\bibfield  {journal} {\bibinfo
  {journal} {ACS Applied Energy Materials}\ }\textbf {\bibinfo {volume} {5}},\
  \bibinfo {pages} {5146} (\bibinfo {year} {2022})}\BibitemShut {NoStop}%
\bibitem [{\citenamefont {Koley}\ \emph {et~al.}(2021)\citenamefont {Koley},
  \citenamefont {Lakshan}, \citenamefont {Raghuvanshi}, \citenamefont {Singh},
  \citenamefont {Bhattacharya},\ and\ \citenamefont
  {Jana}}]{koley2021ultralow}%
  \BibitemOpen
  \bibfield  {author} {\bibinfo {author} {\bibfnamefont {B.}~\bibnamefont
  {Koley}}, \bibinfo {author} {\bibfnamefont {A.}~\bibnamefont {Lakshan}},
  \bibinfo {author} {\bibfnamefont {P.~R.}\ \bibnamefont {Raghuvanshi}},
  \bibinfo {author} {\bibfnamefont {C.}~\bibnamefont {Singh}}, \bibinfo
  {author} {\bibfnamefont {A.}~\bibnamefont {Bhattacharya}},\ and\ \bibinfo
  {author} {\bibfnamefont {P.~P.}\ \bibnamefont {Jana}},\ }\bibfield  {title}
  {\bibinfo {title} {Ultralow lattice thermal conductivity at room temperature
  in cu4tise4},\ }\href {https://doi.org/10.1002/ange.202014222} {\bibfield
  {journal} {\bibinfo  {journal} {Angewandte Chemie}\ }\textbf {\bibinfo
  {volume} {133}},\ \bibinfo {pages} {9188} (\bibinfo {year}
  {2021})}\BibitemShut {NoStop}%
\bibitem [{\citenamefont {Fallah}\ and\ \citenamefont
  {Moghaddam}(2021)}]{fallah2021ultra}%
  \BibitemOpen
  \bibfield  {author} {\bibinfo {author} {\bibfnamefont {M.}~\bibnamefont
  {Fallah}}\ and\ \bibinfo {author} {\bibfnamefont {H.~M.}\ \bibnamefont
  {Moghaddam}},\ }\bibfield  {title} {\bibinfo {title} {Ultra-low lattice
  thermal conductivity and high thermoelectric efficiency in cs2snx6 (x= br,
  i): A dft study},\ }\href {https://doi.org/10.1016/j.mssp.2021.105984}
  {\bibfield  {journal} {\bibinfo  {journal} {Materials Science in
  Semiconductor Processing}\ }\textbf {\bibinfo {volume} {133}},\ \bibinfo
  {pages} {105984} (\bibinfo {year} {2021})}\BibitemShut {NoStop}%
\bibitem [{\citenamefont {Gibson}\ \emph {et~al.}(2021)\citenamefont {Gibson},
  \citenamefont {Zhao}, \citenamefont {Daniels}, \citenamefont {Walker},
  \citenamefont {Daou}, \citenamefont {H{\'e}bert}, \citenamefont {Zanella},
  \citenamefont {Dyer}, \citenamefont {Claridge}, \citenamefont {Slater} \emph
  {et~al.}}]{gibson2021low}%
  \BibitemOpen
  \bibfield  {author} {\bibinfo {author} {\bibfnamefont {Q.~D.}\ \bibnamefont
  {Gibson}}, \bibinfo {author} {\bibfnamefont {T.}~\bibnamefont {Zhao}},
  \bibinfo {author} {\bibfnamefont {L.~M.}\ \bibnamefont {Daniels}}, \bibinfo
  {author} {\bibfnamefont {H.~C.}\ \bibnamefont {Walker}}, \bibinfo {author}
  {\bibfnamefont {R.}~\bibnamefont {Daou}}, \bibinfo {author} {\bibfnamefont
  {S.}~\bibnamefont {H{\'e}bert}}, \bibinfo {author} {\bibfnamefont
  {M.}~\bibnamefont {Zanella}}, \bibinfo {author} {\bibfnamefont {M.~S.}\
  \bibnamefont {Dyer}}, \bibinfo {author} {\bibfnamefont {J.~B.}\ \bibnamefont
  {Claridge}}, \bibinfo {author} {\bibfnamefont {B.}~\bibnamefont {Slater}},
  \emph {et~al.},\ }\bibfield  {title} {\bibinfo {title} {Low thermal
  conductivity in a modular inorganic material with bonding anisotropy and
  mismatch},\ }\href {https://doi.org/10.1126/science.abh1619} {\bibfield
  {journal} {\bibinfo  {journal} {Science}\ }\textbf {\bibinfo {volume}
  {373}},\ \bibinfo {pages} {1017} (\bibinfo {year} {2021})}\BibitemShut
  {NoStop}%
\bibitem [{\citenamefont {Zhang}\ \emph {et~al.}(2023)\citenamefont {Zhang},
  \citenamefont {Yu}, \citenamefont {Ning}, \citenamefont {Zhang},
  \citenamefont {Qi}, \citenamefont {Jiang},\ and\ \citenamefont
  {Chen}}]{zhang2023extremely}%
  \BibitemOpen
  \bibfield  {author} {\bibinfo {author} {\bibfnamefont {T.}~\bibnamefont
  {Zhang}}, \bibinfo {author} {\bibfnamefont {T.}~\bibnamefont {Yu}}, \bibinfo
  {author} {\bibfnamefont {S.}~\bibnamefont {Ning}}, \bibinfo {author}
  {\bibfnamefont {Z.}~\bibnamefont {Zhang}}, \bibinfo {author} {\bibfnamefont
  {N.}~\bibnamefont {Qi}}, \bibinfo {author} {\bibfnamefont {M.}~\bibnamefont
  {Jiang}},\ and\ \bibinfo {author} {\bibfnamefont {Z.}~\bibnamefont {Chen}},\
  }\bibfield  {title} {\bibinfo {title} {Extremely low lattice thermal
  conductivity leading to superior thermoelectric performance in cu4tise4},\
  }\href {https://doi.org/10.1021/acsami.3c05602} {\bibfield  {journal}
  {\bibinfo  {journal} {ACS Applied Materials \& Interfaces}\ }\textbf
  {\bibinfo {volume} {15}},\ \bibinfo {pages} {32453} (\bibinfo {year}
  {2023})}\BibitemShut {NoStop}%
\bibitem [{\citenamefont {Cutler}\ \emph {et~al.}(1964)\citenamefont {Cutler},
  \citenamefont {Leavy},\ and\ \citenamefont
  {Fitzpatrick}}]{cutler1964electronic}%
  \BibitemOpen
  \bibfield  {author} {\bibinfo {author} {\bibfnamefont {M.}~\bibnamefont
  {Cutler}}, \bibinfo {author} {\bibfnamefont {J.}~\bibnamefont {Leavy}},\ and\
  \bibinfo {author} {\bibfnamefont {R.}~\bibnamefont {Fitzpatrick}},\
  }\bibfield  {title} {\bibinfo {title} {Electronic transport in semimetallic
  cerium sulfide},\ }\href {https://doi.org/10.1103/PhysRev.133.A1143}
  {\bibfield  {journal} {\bibinfo  {journal} {Physical Review}\ }\textbf
  {\bibinfo {volume} {133}},\ \bibinfo {pages} {A1143} (\bibinfo {year}
  {1964})}\BibitemShut {NoStop}%
\bibitem [{\citenamefont {Snyder}\ and\ \citenamefont
  {Toberer}(2008)}]{snyder2008complex}%
  \BibitemOpen
  \bibfield  {author} {\bibinfo {author} {\bibfnamefont {G.~J.}\ \bibnamefont
  {Snyder}}\ and\ \bibinfo {author} {\bibfnamefont {E.~S.}\ \bibnamefont
  {Toberer}},\ }\bibfield  {title} {\bibinfo {title} {Complex thermoelectric
  materials},\ }\href {https://doi.org/10.1038/nmat2090} {\bibfield  {journal}
  {\bibinfo  {journal} {Nature Materials}\ }\textbf {\bibinfo {volume} {7}},\
  \bibinfo {pages} {105} (\bibinfo {year} {2008})}\BibitemShut {NoStop}%
\bibitem [{\citenamefont {Glassbrenner}\ and\ \citenamefont
  {Slack}(1964)}]{glassbrenner1964thermal}%
  \BibitemOpen
  \bibfield  {author} {\bibinfo {author} {\bibfnamefont {C.~J.}\ \bibnamefont
  {Glassbrenner}}\ and\ \bibinfo {author} {\bibfnamefont {G.~A.}\ \bibnamefont
  {Slack}},\ }\bibfield  {title} {\bibinfo {title} {{Thermal conductivity of
  silicon and germanium from 3 K to the melting point}},\ }\href
  {https://doi.org/10.1103/PhysRev.134.A1058} {\bibfield  {journal} {\bibinfo
  {journal} {Physical Review}\ }\textbf {\bibinfo {volume} {134}},\ \bibinfo
  {pages} {A1058} (\bibinfo {year} {1964})}\BibitemShut {NoStop}%
\bibitem [{\citenamefont {Shi}\ \emph {et~al.}(2015)\citenamefont {Shi},
  \citenamefont {Parker}, \citenamefont {Du},\ and\ \citenamefont
  {Singh}}]{shi2015connecting}%
  \BibitemOpen
  \bibfield  {author} {\bibinfo {author} {\bibfnamefont {H.}~\bibnamefont
  {Shi}}, \bibinfo {author} {\bibfnamefont {D.}~\bibnamefont {Parker}},
  \bibinfo {author} {\bibfnamefont {M.-H.}\ \bibnamefont {Du}},\ and\ \bibinfo
  {author} {\bibfnamefont {D.~J.}\ \bibnamefont {Singh}},\ }\bibfield  {title}
  {\bibinfo {title} {{Connecting thermoelectric performance and
  topological-insulator behavior: Bi2Te3 and Bi2Te2Se from first principles}},\
  }\href {https://doi.org/10.1103/PhysRevApplied.3.014004} {\bibfield
  {journal} {\bibinfo  {journal} {Physical Review Applied}\ }\textbf {\bibinfo
  {volume} {3}},\ \bibinfo {pages} {014004} (\bibinfo {year}
  {2015})}\BibitemShut {NoStop}%
\bibitem [{\citenamefont {Gong}\ \emph {et~al.}(2016)\citenamefont {Gong},
  \citenamefont {Hong}, \citenamefont {Shuai}, \citenamefont {Li},
  \citenamefont {Yan}, \citenamefont {Ren},\ and\ \citenamefont
  {Liu}}]{gong2016investigation}%
  \BibitemOpen
  \bibfield  {author} {\bibinfo {author} {\bibfnamefont {J.}~\bibnamefont
  {Gong}}, \bibinfo {author} {\bibfnamefont {A.}~\bibnamefont {Hong}}, \bibinfo
  {author} {\bibfnamefont {J.}~\bibnamefont {Shuai}}, \bibinfo {author}
  {\bibfnamefont {L.}~\bibnamefont {Li}}, \bibinfo {author} {\bibfnamefont
  {Z.}~\bibnamefont {Yan}}, \bibinfo {author} {\bibfnamefont {Z.}~\bibnamefont
  {Ren}},\ and\ \bibinfo {author} {\bibfnamefont {J.-M.}\ \bibnamefont {Liu}},\
  }\bibfield  {title} {\bibinfo {title} {{Investigation of the bipolar effect
  in the thermoelectric material CaMg2Bi2 using a first-principles study}},\
  }\href {https://doi.org/10.1039/C6CP02057G} {\bibfield  {journal} {\bibinfo
  {journal} {Physical Chemistry Chemical Physics}\ }\textbf {\bibinfo {volume}
  {18}},\ \bibinfo {pages} {16566} (\bibinfo {year} {2016})}\BibitemShut
  {NoStop}%
\end{thebibliography}%

\end{document}